\documentclass[%
 reprint,
 floatfix,
groupedaddress,
longbibliography,
 amsmath,amssymb,
 aps,
 prx,
]{revtex4-2}
\makeatletter
\def\set@curr@file#1{%
 \begingroup
 \escapechar\m@ne
 \xdef\@curr@file{\expandafter\string\csname #1\endcsname}%
 \endgroup
}
\def\quote@name#1{"\quote@@name#1\@gobble""}
\def\quote@@name#1"{#1\quote@@name}
\def\unquote@name#1{\quote@@name#1\@gobble"}
\makeatother
\usepackage{graphicx}
\usepackage{dcolumn}
\usepackage{bm}
\usepackage{xfrac}
\usepackage{siunitx}
\usepackage{amsmath}
\DeclareMathOperator{\sign}{sign}
\usepackage[utf8]{inputenc}
\usepackage[T1]{fontenc}
\usepackage[export]{adjustbox}
\usepackage[colorinlistoftodos,prependcaption,textsize=tiny]{todonotes}
\usepackage{gensymb}
\presetkeys%
 {todonotes}%
 {inline, backgroundcolor=yellow}{}

\begin{document}
\preprint{APS/123-QED}

\title{Boiling in Nanopores through Localized Joule Heating:\\Transition between Nucleate and Film Boiling}

\author{Soumyadeep Paul}
\affiliation{%
Department of Mechanical Engineering, The University of Tokyo, 7-3-1, Hongo, Bunkyo-ku, Tokyo 113-8656, Japan
}%
\author{Wei-Lun Hsu}%
\affiliation{%
Department of Mechanical Engineering, The University of Tokyo, 7-3-1, Hongo, Bunkyo-ku, Tokyo 113-8656, Japan
}%
\author{Yusuke Ito}%
\affiliation{%
Department of Mechanical Engineering, The University of Tokyo, 7-3-1, Hongo, Bunkyo-ku, Tokyo 113-8656, Japan
}%
\author{Hirofumi Daiguji}%
 \email{Corresponding author: daiguji@thml.t.u-tokyo.ac.jp}
\affiliation{%
Department of Mechanical Engineering, The University of Tokyo, 7-3-1, Hongo, Bunkyo-ku, Tokyo 113-8656, Japan
}%





\begin{abstract}
The transition from nucleate to film boiling on micro/nano textured surfaces is of crucial importance in a number of practical applications, where it  needs to be avoided to enable safe and efficient heat transfer. Previous studies have focused on the transition process at the macroscale, where heat transfer and bubble generation are activated on an array of micro/nanostructures. In the present study, we narrow down our investigation scale to a single nanopore, where, through localized Joule heating within the pore volume, single-bubble nucleation and transition are examined at nanosecond resolution using resistive pulse sensing and acoustic sensing. Akin to macroscale boiling, where heterogeneous bubbles can nucleate and coalesce into a film, in the case of nanopores also, patches of heterogeneous bubbles nucleating on the cylindrical pore surface can form a torus-shaped vapor film blanketing the entire pore surface. In contrast to conventional pool boiling, nanopore boiling involves a reverse transition mechanism, where, with increased heat generation, film boiling reverts  to nucleate boiling. With increasing bias voltage across the nanopore, the Joule heat production increases within the pore, leading to destabilization and collapse of the torus-shaped vapor film.

\begin{description}
\item[Keywords]
Nanopore, Ionic Joule heating, Bubble nucleation, Boiling transitions,\\ Resistive pulse sensing, Acoustic sensing, Torus bubble
\end{description}
\end{abstract}

\maketitle


\section{INTRODUCTION}\label{sec:level1}
Despite a century of research on vapor bubble dynamics and its ramifications for boiling phenomena, a unified understanding of bubble nucleation at the nanoscale and growth/transition to macrobubbles has yet to be established. Little attention has been paid to establishing fundamental connections between boiling characteristics at the nanoscale and macroscale, which may differ significantly owing to confinement effects~\cite{hill2001different, bocquet2020nanofluidics}. Closing this fundamental gap in knowledge is of paramount importance for fully leveraging the benefits of phase change heat transfer in high-heat-flux applications, including, among others, electronic cooling~\cite{Dhillon2015}, inkjet printing~\cite{Asai1989}, and spray quenching~\cite{jiang2022inhibiting}. Starting with the work of Nukiyama~\cite{nukiyama1984memories}, who heated a liquid by running electricity through a metal wire, early research in this field was devoted to establishing a boiling curve consisting of five regions of pool boiling~\cite{lienhard1985historical}: natural convection, isolated nucleate boiling followed by slug nucleate boiling, transition boiling terminating at the Leidenfrost point, and ultimately film boiling. 
Research has been focused on fundamental understanding of pool boiling transitions with the aim of achieving practical goals in engineering~\cite{Dhillon2015, Chen2006, Kandlikar2007}, such as  enhancement of the heat transfer coefficient~\cite{cho2015turning} and prevention of early critical heat flux~\cite{Dhillon2015}. The latter is the limiting heat flux beyond which the heat transfer to the liquid starts to decrease owing to lack of liquid contact with the heated surface because of horizontal coalescence of nucleating bubbles.

From a fundamental viewpoint, to understand the initial stage of boiling, it is imperative to investigate the dynamics of single vapor nanobubbles. This is especially challenging for two reasons: (i) heating must be focused such that single-bubble nucleation can be achieved; (ii)  the nanoscale--nanosecond dynamics of the nanobubble post nucleation must be captured experimentally. Because of these inherent difficulties, at the nanoscale,   engineering research on boiling involving heat transfer enhancement~\cite{Dhir2013} and physics research on single-vapor bubble dynamics involving equilibrium and stability characteristics~\cite{Prosperetti2017} have mostly proceeded in parallel but separately. On the other hand, at the macroscale,  engineering studies of boiling and of  single-vapor-bubble dynamics have been performed simultaneously. In this regard, we would highlight the work of Dhir and co-workers~\cite{qiu2002single} in which microgravity was used to confine a macrobubble on a heater surface, thereby allowing a prolonged investigation of spherical thermal bubble dynamics. 

At the outset, we formulate  three critical questions that need to be addressed to bridge the gap between engineering boiling research and physics research on single-vapor bubble dynamics at the nanoscale: (i) How do bubble growth and transition dynamics at the nanoscale differ from those at the macroscale? (ii) What is the connection between the governing dynamics at the two scales? (iii) How can vapor bubble seeds improve the engineering limits of boiling heat transfer. Both from the perspective of thermodynamics~\cite{hill2001different} and that of fluid mechanics~\cite{bocquet2020nanofluidics}, the governing dynamics at the 100~nm scale are expected be different owing to confinement effects, leading to different bifurcation mechanisms of the boiling structure~\cite{chai2001boiling, chai2002self, shoji2004studies}. Owing to the lack of sufficient experimental studies and theoretical understanding of nanoscale boiling, we first need to address the first question and clarify boiling at the single-nanobubble limit, before the scale gap between nanoscale and macroscale boiling can be bridged.

\begin{figure*}[!p]
\includegraphics[width=\textwidth,keepaspectratio,angle=0]{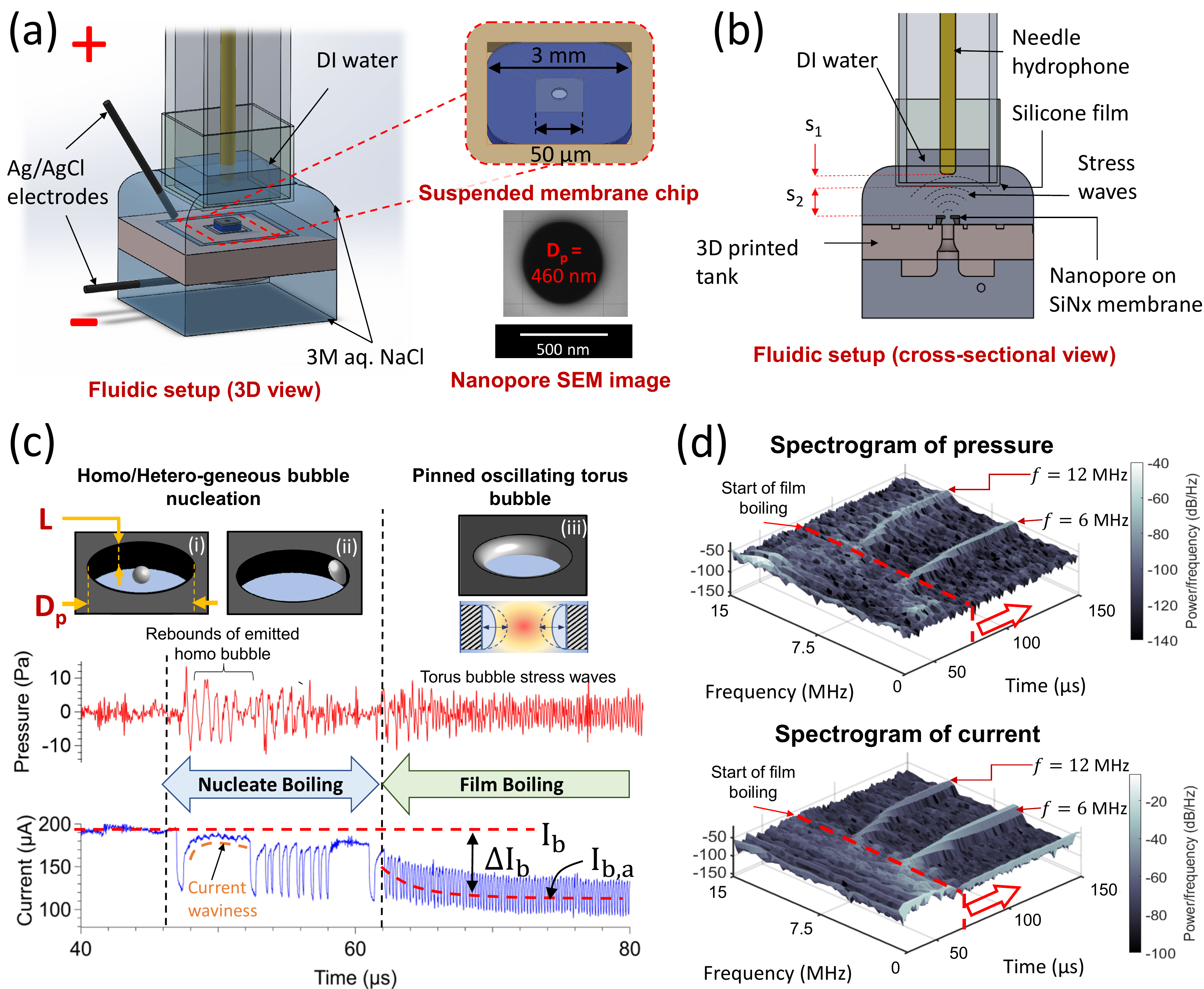}
\caption{\label{fig:1} (a)  3D schematic view and (b) cross-sectional schematic view of the acoustic and resistive pulse sensing experimental setup. The insets in (a) show a schematic of the suspended membrane nanopore chip (a cylindrical hole of diameter $D_\mathrm{p}$ on a silicon nitride thin layer of thickness $L$) and an SEM image of the $D_\mathrm{p}= 460$~nm nanopore. (c) Nanopore current (blue) and hydrophone pressure (red) traces during nanopore boiling during a \SI{1468}{\micro\second}--7~V voltage pulse starting from $t=$~\SI{0}{\micro\second}. Before \SI{46}{\micro\second}, when there is only superheating and no boiling, both the current and pressure baselines are steady. This is followed by stochastic nucleate boiling, when two processes of bubble generation are predominant: (i) homogeneous bubble nucleation at the pore center and (ii) patch heterogeneous bubble nucleation on the pore surface. Here, discrete homogeneous and heterogeneous bubbles grow and shrink, nonperiodic current blockage signals are seen in the current trace, and high-amplitude nonperiodic pressure waves are observed. When nucleate boiling transitions to film boiling after $\sim$\SI{16}{\micro\second}, a torus-shaped vapor bubble blankets the nanopore wall surface (iii). This bubble then performs pinned volumetric oscillations in thermal resonance with the Joule heat generation inside the pore liquid. Owing to the stability of the torus bubble of  volume comparable  to the pore volume $V_\mathrm{p}$, the baseline current steadily decreases by $\sim42\%$ from $I_\mathrm{b}$ to $I_\mathrm{b,a}$. The large shift in current baseline ($\Delta I_\mathrm{b}$) for a pore with a  high $D_\mathrm{p}/L$ ratio  ($=4.6$) can only be explained by the existence of a torus bubble that has a volume $V_\mathrm{b}$ scaling proportionally with the pore diameter. Owing to the high sensitivity of the nanopore current to the insulating bubble, oscillations of the bubble volume about its mean value are represented by near-sinusoidal current oscillations about the reduced baseline $I_\mathrm{b,a}$. Consequently, the hydrophone also picks up high-frequency and periodic pressure waves generated as a result of bubble volume oscillations. (d) 3D spectrogram analysis of the nanopore current and hydrophone pressure. During nucleate boiling with stochastic waiting times, the power spectrum is distributed across a wide frequency band, and hence no clear frequency ridges are seen in the spectrograms. After the film boiling transition at $\sim$\SI{62}{\micro\second}, when the torus bubble oscillates at a characteristic frequency, two ridges are seen at $f=6$~MHz and $12$~MHz,  signifying the first and second harmonics, respectively.}
\end{figure*}

To activate single-bubble nucleation, we utilize Joule heating, which directly converts electric energy to thermal energy in confined liquid volumes, circumventing interfacial heat transfer. Nagashima \emph{et al.}~\cite{Nagashima2014} filled a concentrated electrolyte solution in a nanopore connected with two large solution reservoirs. An electric potential bias was imposed across a membrane of thickness 71~nm  via electrodes inserted in each side of the reservoir. The electric field was focused around the nano-aperture on the thin membrane, generating intense Joule heating in the nanospace. Homogeneous thermal bubble generation in the liquid phase was successfully detected via ionic current measurements in the gigahertz bandwidth of resistive pulses, based on the volume exclusion effect of nonconductive bubbles. This method of bubble generation and detection overcomes both the scale and time limitations of traditional boiling studies using heater surfaces and high-speed cameras, thereby serving as an ideal platform for tracking single nanobubbles in the initial stage of boiling. Using the same platform, Paul \emph{et al.}~\cite{paul2020single} demonstrated that solely homogeneous nucleation  occurs only in tiny apertures at the nanoscale. As the pore diameter expands, heterogeneous bubbles originating from the inner walls of nanopores become dominant. This transition was attributed  both to geometric confinement effects such as contact line pinning and to a thermal confinement effect, namely, a large temperature gradient within the nanopore.

In the present paper, we measure both ionic current variations and stress waves created by a nanopore bubble [Figs.~\ref{fig:1}(a) and~\ref{fig:1}(b)], focusing specifically on the transition between nucleate and film boiling. By intensifying the heating rate through increasing the bias voltages across the nanopore, we activate the transition from nucleate boiling to heterogeneous film boiling on the rim of the nanopore. Owing to the cylindrical shape of the pore, the film bubble resembles a torus in shape, as shown in Fig.~\ref{fig:1}(c) [Appendix~\ref{appA}]. A similar observation was also observed in Medvedev \emph{et al.}~\cite{medvedev2013calculations}, albeit at macroscale. Nanobubbles on flat surfaces have already been shown to exhibit remarkable stability owing to the contact line pinning effect~\cite{Lohse2015, nag2021dynamic}, and here we  demonstrate that nano-torus thermal bubbles also exhibit stability owing to similar pining effects within nanopores. Unlike nucleate bubbles, which undergo rapid growth and collapse cycles separated by significant reheating periods, the film bubble is stable and relatively long-lived, undergoing pinned volumetric oscillations, which are captured by our sensors in the form of pore current and hydrophone pressure oscillations. When the bubble is moderately stable, the oscillations are weakly nonlinear~\cite{brennen2014cavitation, lauterborn2010physics} leading to frequency dispersion, and only the fundamental and second harmonics are observed beyond the noise level of our sensors. The 3D spectrogram analysis of the torus bubble presented in Fig.~\ref{fig:1}(d)   reveals two ridges at 6~MHz and 12~MHz which correspond to these two harmonics. These frequencies are one order higher than those recorded for oscillatory boiling on microheaters~\cite{Li2017a}, probably due to the larger bubble size compared to the nanopore bubble. 
Intriguingly, our voltammetric studies reveal that for larger pore diameters, with increasing voltage, the torus bubble gradually loses its stability, ultimately leading to transition from film to nucleate boiling.

In this paper, we showcase the similarities in bubble dynamics and thermodynamics between macroscale pool boiling and nanopore boiling, while also highlighting the differences in the thermofluidic mechanisms driving the transitions. The remainder of the paper is organized as follows. In Sec.~\ref{secII}, a brief overview of our experimental system is presented. Section~\ref{secIIIA} explains the nucleate-to-film boiling transition using current and hydrophone pressure spectrograms for a 420~nm pore, and Sec.~\ref{secIIIB} elucidates the film-to~nucleate boiling transition in a 460~nm pore. Section~\ref{secIVA} discusses the effect on Joule heat generation during the different nanopore boiling regimes for the two pore sizes. In macroscale film boiling, the vapor film limits heat \textit{transfer} from the heated surface, whereas in nanopore boiling, the torus vapor film limits Joule heat \textit{production} through the volume exclusion effect. In Sec.~\ref{secIVB}, a theoretical model is developed to capture the equilibrium size and temperature of the torus vapor film at different applied voltages. In Sec.~\ref{secIVC}, the stability of the torus vapor film is analyzed by perturbation theory, and the reverse transition phenomenon (film-to-nucleate boiling) is explained based on this analysis. Conclusions are presented in Sec.~\ref{secV}/

\section{EXPERIMENTAL METHODS}\label{secII}
Circular nanopores as shown in the scanning electron microscope (SEM) image  in Fig.~\ref{fig:1}(a) and in Fig.~S1 of the Supplemental Material~\cite{supp} were made on silicon nitride chips (Model No.~4088SN-BA) purchased from Alliance Biosystems Inc., each comprising a 100-nm-thick silicon nitride (Si$_{3}$N$_{4}$) membrane deposited on a 200-µm-thick silicon substrate with an approximately square $\SI{50}{\micro\meter} \times \SI{50}{\micro\meter}$ opening at the center. The nanopores were etched at the center of the free-standing part of the membrane using a focused $\text{Ga}^+$ ion beam (SMI2050MS2, SII Nanotechnology). Post fabrication, the nanopore chip was assembled between two fluid tanks and wetted with ethanol before flushing with a 3M aqueous solution of NaCl prepared by diluting 5M NaCl solution (Sigma-Aldrich) with deionized (DI) water. Voltage pulses were applied through Ag/AgCl electrodes using a  pulse generator (Tektronix AFG3152C), and the ionic current flowing through the nanopore was registered on a  oscilloscope (Tektronix MSO56) by measuring the voltage across a shunt resistor through an active power rail probe at 20~MHz terminal bandwidth (Tektronix  TPR4000). In addition, a passive probe set at 250~MHz terminal bandwidth was used to measure high-frequency current oscillations across a second shunt resistor in series. The active probe, which had a high signal-to-noise ratio but also high capacitance, was used to track the baseline and frequency shifts of the nanopore current, while the passive probe measurements taken at high bandwidth were used to measure the amplitude of current oscillations.  A needle hydrophone (sensor diameter 4~mm, Precision Acoustics NH4000) encapsulated in a hollow quartz cell was also placed vertically above the chip surface to collect the stress waves from the nanopore bubbles at 20~MHz terminal bandwidth. The stress waves collected by the piezoelectric sensor were converted into electrical signals by a pre-amplifier and DC coupler, which were registered in the oscilloscope concurrently with the current signals. To shield the piezoelectric element of the hydrophone from the ionic current, the hydrophone was encapsulated in a custom-made glass shell filled with DI water. A 20-µm-thick silicone film (Wacker Asahikasei Silicone) [Fig.~\ref{fig:1}(b)] separated the DI water from the salt solution, allowing acoustic signals to pass. Hydrophobic tape (3M Microfluidic Diagnostic Tape 9965) was used to seal the junction of the silicone film and quartz cell to prevent any electrolyte leakage. As a result of hydrophone encapsulation, the clearance distance between the piezoelectric sensor and the nanopore ($s=s_1+s_2$) had two components, $s_1$ and $s_2$ [Fig.~\ref{fig:1}(b)]. The distance between sensor and silicone film, $s_1$, was measured by an optical microscope, while $s_2$ was measured first using a contact-type distance sensor and this measurement was validated based on the delay time between the current dip and the hydrophone peak signal for homogeneous bubble nucleation (see Sec.~S2 in the Supplemental Material~\cite{supp}). To measure the low-intensity stress waves emanating from torus bubble oscillations at an acceptable signal-to-noise ratio, the net clearance needed to be as small as possible. For the 460~nm pore experiments detailed in Figs.~\ref{fig:1} and~\ref{fig:3}, $s=500\pm20$~µm. Owing to the large difference between sensor diameter and clearance distance, the spatial averaging effect~\cite{radulescu2001hydrophone} distorted the actual stress wave amplitude. Nonetheless, a qualitative analysis of pressure amplitudes for varying voltages and varying bubble frequencies still provides insight into the bubble dynamics.

The acoustic signals registered in the oscilloscope as voltage signals $V_\mathrm{p}(t)$ were then converted into pressure waveforms using the following equation~\cite{hurrell2004voltage}:
\begin{equation}
p(t) =  F^{-1}\left\{\frac{F\{V_\mathrm{p}(t)\}}{M(f)}\right \},
\label{eq:ap1}
\end{equation}
where  $F$ and the $F^{-1}$ operators denote the Fourier and inverse Fourier transforms, respectively, and $M(f)$ is the frequency response of the hydrophone sensitivity as per plane wave calibration measurements performed by the manufacturer, Precision Acoustics. The uncertainty in $M(f)$ was in the range of 19--22\%. The imaginary part of the acoustic pressure $p(t)$ obtained after inverse Fourier transform is neglected.

\section{Results}\label{secIII}
Single-bubble boiling is activated by localized Joule heating inside a submicrometer pore on a 100-nm-thick suspended silicon nitride membrane [Fig.~\ref{fig:1}(a)]. The nanopore is submerged inside a 3M NaCl electrolyte solution, which leads to ionic current flow once bias voltages are applied across it through Ag/AgCl electrodes. When homo- or heterogeneous bubbles are nucleated beyond their respective superheating limits, the ionic current is altered, and these changes are  measured by a high-bandwidth oscilloscope. In addition, an encapsulated piezoelectric hydrophone  on top of the nanopore chip [Fig.~\ref{fig:1}(b)] absorbs the stress waves generated by bubble motion and converts them into an electrical signal to be recorded in the oscilloscope simultaneously. By studying the amplitude and frequency variations of current and hydrophone pressure, we can track the boiling transition within the pore. The current signals are most sensitive to bubble dynamics within the pore, where minor changes in bubble volume modulate the pore cross-section significantly. Meanwhile, the hydrophone sensor catches any stress waves in the liquid generated by bubble expansion or shrinkage, irrespective of the bubble position relative to the pore.

\subsection{Nucleate-to-film boiling transition}\label{secIIIA}
\begin{figure*}[!t]
\includegraphics[width=\textwidth,keepaspectratio,angle=0]{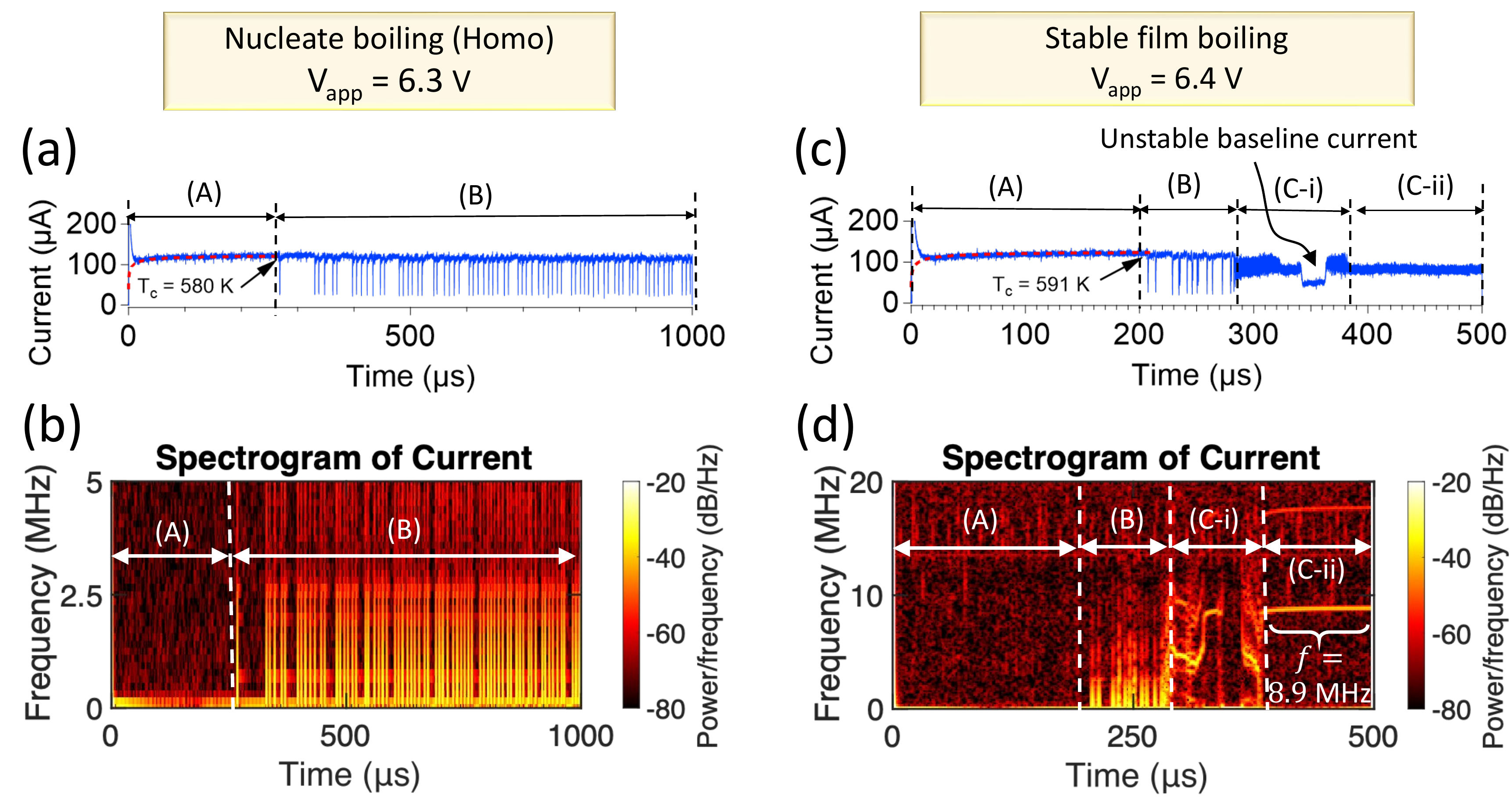}
\caption{\label{fig:2} Boiling transition with increasing bias voltage for a $D_\mathrm{p}= 420$~nm pore. (a) and (b)  Nucleate homogeneous boiling at 6.3~V. (c) and (d) Transition to stable film boiling at 6.4~V. (a) and (c) are plots of the transient shift in baseline current during the voltage pulse, while (b) and (d) are spectrograms of the current corresponding to the bubble oscillation frequency.}
\end{figure*}

Figure~\ref{fig:2} shows the boiling structure of a $D_\mathrm{p}=420$~nm pore. The current--time traces in (a) and (c) show baseline current changes and spectrograms (short time averaged Fast Fourier Transform) in (b) and (d) show frequency changes, both providing phenomenological evidence of nucleate to film boiling transition.

When a voltage pulse of 6--8~V is triggered, ionic current flowing through the nanopore liberates Joule heat~\cite{Viasnoff2010, tsutsui2022ionic}, which transiently increases the nanopore temperature at a rate of $\sim\!10^7$~K/s from ambient conditions of 298~K~\cite{paul2021analysis}. As the temperature rises, the ionic conductivity increases, allowing more current flow and Joule heat generation in a feedback loop. Unlike pool boiling, where the substrate temperature can be controlled directly, in this case modulating the bias voltage only allows us to control the heat generation rate, which can reach $\sim\!10^{16}$~\SI[per-mode = symbol]{}{\watt\per\metre^3}~\cite{paul2021analysis, paul2022data}. Owing to the rate of high heating  within the confined space, a temperature gradient of the order of 1~K/nm  develops within the pore, which allows bubbles of different volumes and temperature to exist in thermal equilibrium within the pore liquid.

The current response to each pulse comprises  an initial heating zone (A) followed by a nucleate boiling zone (B), ultimately leading  to torus film boiling (C). At 6.3~V, except for an outlier (see Fig.~S17 in the Supplemental Material~\cite{supp}),  only  zones A and B are present during boiling [Figs.~\ref{fig:2}(a) and~\ref{fig:2}(b)]. However, with a 6.4~V voltage pulse, all  three zones can be clearly identified, as shown in Fig.~\ref{fig:2}(c) and~\ref{fig:2}(d). This scheme is seen for multiple voltage pulses of the same magnitude. Thus,  there is an increasing probability of film boiling transition  as the voltage is increased from 6.3~V to 6.4~V.

In zone A, the Joule heat generation within the pore is balanced by heat dissipation by the silicon nitride membrane and surrounding electrolyte. No bubble-induced current blockage signals are seen, and the spectrograms also show no effect [Figs.~\ref{fig:2}(a) and~\ref{fig:2}(b)]. Zone A terminates with  nucleation of a homogeneous bubble at the pore center when the local temperature reaches $T_\mathrm{c}\sim590\pm10$~K. This temperature estimate  was obtained through numerical simulations, by fitting the experimental nanopore current [the red trace in Fig.~\ref{fig:2}(a)] as described in Appendix~\ref{appB}. Before this bubble nucleation, although the pore surface temperature matches the patch heterogeneous nucleation temperature of $472\pm35$~K~\cite{Witharana2012}, these bubbles are suppressed. Our previous paper~\cite{paul2020single} on this topic showed that during Joule heating, unstable vapor clusters can form homogeneously at the pore center with temperature $T_\mathrm{c}$ and  heterogeneously on the pore surface with temperature  $T_\mathrm{w}$. Depending on the value of the cross-pore temperature difference $\Delta T_\mathrm{p}=T_\mathrm{c} - T_\mathrm{w}$, a cluster ripening competition is established between these two cluster groups at the two nucleation sites. When $\Delta T_\mathrm{p}$ is higher,  homogeneous cluster growth requires less free energy to grow, and hence the heterogeneous clusters are suppressed.

Figure~\ref{fig:2}(a) shows only the nucleate boiling regime (B) involving quasiperiodic homogeneous bubble formation and ejection at 6.3~V for a 420~nm pore. It should be noted that unlike  patch bubbles, which grow and collapse in a pinned state~\cite{Lohse2015, paul2020single, zou2018surface} on the pore surface, the homogeneous bubbles are ejected from the pore by the electric field force acting on its negatively charged surface (the typical surface charge is \SI{-23}{\milli\coulomb\per\meter^2}~\cite{Hu2018}). The homogeneous bubble is retained near the pore access region, where it rebounds spherically and volumetrically, emitting high-amplitude stress waves. Additionally, owing to the limited influence of bubble volume fluctuations on the ion current in the pore access region, low-amplitude current waviness is seen in the reheating current trace [Fig.~\ref{fig:1}(c)] after the first growth cycle post nucleation~\cite{lauterborn2010physics}. This phenomenon is illustrated in Fig.~S2 of the Supplemental Material~\cite{supp}. This is markedly different from pool boiling, where the departure of the bubble from the nucleation site is mainly buoyancy-driven.
\begin{figure*}[!t]
\includegraphics[width=\textwidth,keepaspectratio,angle=0]{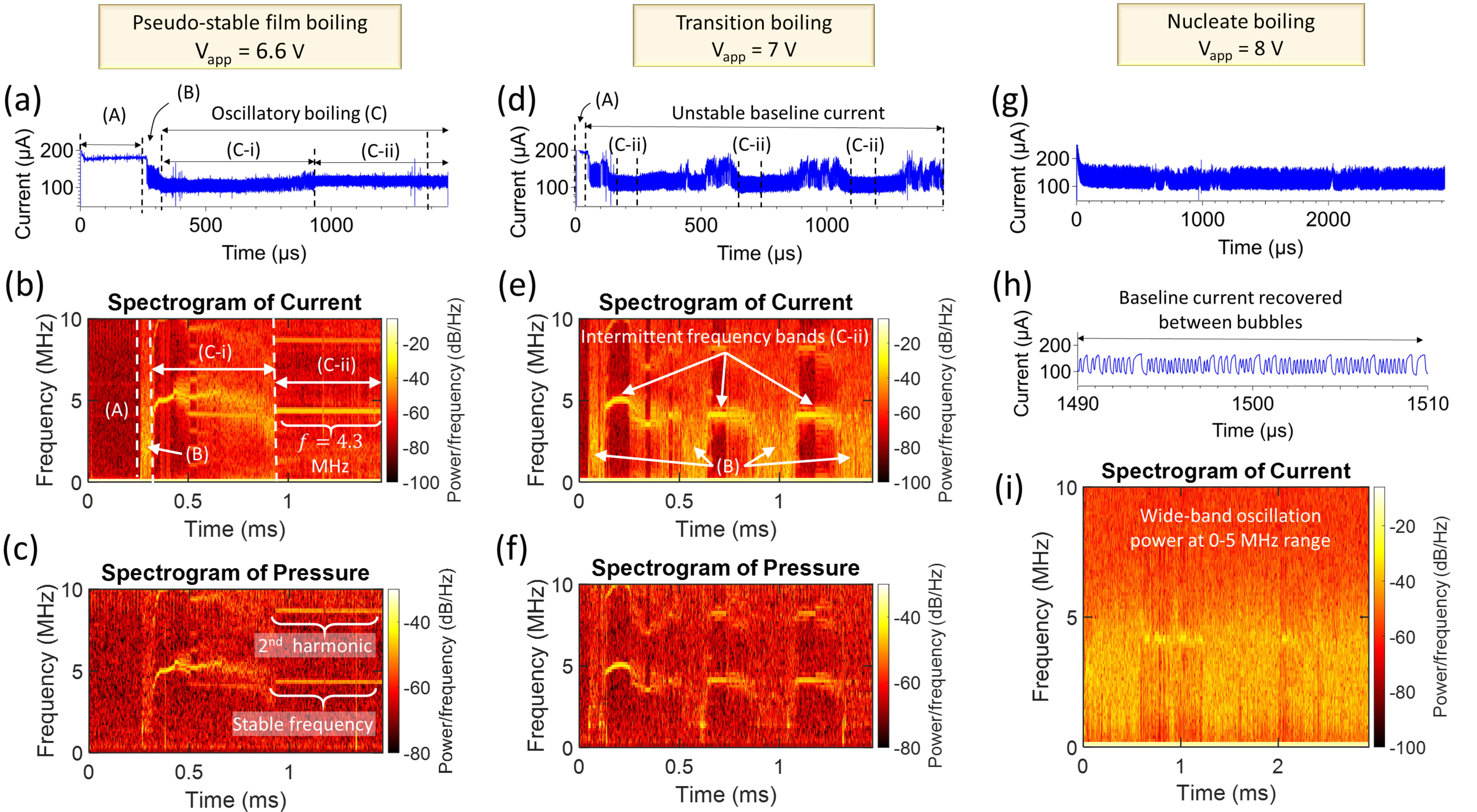}
\caption{\label{fig:3} Boiling transition with increasing bias voltage for a $D_\mathrm{p}= 460$~nm pore. (a) Transient current, (b) current spectrogram, and (c) hydrophone pressure spectrogram during stable oscillatory film boiling at 6.6~V. (d) Transient current, (e) current spectrogram, and (f) hydrophone pressure spectrogram during transition boiling at 7~V. (g) Transient current of the whole voltage pulse and (h) a zoomed view during nucleate boiling at 8~V. (i) Current spectrogram at 8~V. Owing to stochastic nucleation, no clear band is seen, but high oscillation power can be seen in the 0--5~MHz range.}
\end{figure*}
The transition from nucleate to film boiling is seen in Fig.~\ref{fig:2}(c). Following zone B, zone C-i starts, during which the baseline current decreases but does not stabilize. In addition, the spectrograms in Fig.~\ref{fig:2}(d) show unstable frequency bands in this zone. Compared with zone B, the frequency bands are discrete but unsteady, indicating that a single unstable torus bubble is oscillating within the  pore. In this regime, the torus bubble is not in thermal equilibrium with the Joule heating, which leads to rapid variations in the mean bubble size, in addition to highly nonlinear volumetric pinned oscillations about the mean size. Eventually, the torus bubble reaches the equilibrium size corresponding to the applied voltage, and the current oscillations stabilize about a stable mean value. This marks the beginning of the stable oscillatory torus boiling regime (C-ii). In this regime, the baseline current stabilizes, and the current and pressure spectrograms show steady and narrow frequency bands. This regime continues until the end of the voltage pulse. For the 420~nm pore [Fig.~\ref{fig:2}(d)], there is a  stable oscillation frequency of 8.9~MHz, whereas for the 460~nm pore [Fig.~\ref{fig:3}(b)], there is a  pseudostable oscillation frequency of 4.3~MHz. Even in this regime, low-power frequency bands at secondary harmonics are also seen, indicating that the oscillations are weakly nonlinear~\cite{brennen2014cavitation}. For the 420~nm pore, we also performed a boiling structure analysis at 6.5~V and 6.75~V. We found that the stable oscillation zone (C-ii) essentially disappeared and was replaced by low-frequency ($\sim$5~MHz) nonlinear oscillations indicative of an unstable torus bubble (C-i). This gradual loss of torus stability with voltage is observed more prominently for the 460~nm pore and is discussed in Sec.~\ref{secIIIB}.

\subsection{Film-to-nucleate boiling transition}\label{secIIIB}
As the torus bubble forms on the pore circumference [Fig.~\ref{fig:1}(c)], its volume scales linearly with the pore diameter, $V_\mathrm{b}\propto D_\mathrm{p}$, while the cylindrical pore volume where the Joule heat is generated scales quadratically,  $V_\mathrm{p}\propto D_\mathrm{p}^2$.  Thus, for torus bubbles forming on the same pore length $L$ [Fig.~\ref{fig:1}(c)], with increasing $D_\mathrm{p}$, more Joule heat will be liberated within the pore volume than can be blocked by the bubble  through the volume exclusion effect. Hence, according to our hypothesis, the torus bubble will be forced to bulge outward, triggering instability and eventually a reverse transition. Through our experiments, we have found this transition to manifest in the pore diameter range of 400--500~nm. We therefore chose the 460~nm pore results to showcase the boiling structure during this process.

First, compared with the 420~nm pore at 6.3~V, which has a \SI{700}{\micro\second} duration of homogeneous nucleate boiling, the 460~nm pore at 6.6~V has a much shorter nucleate boiling zone of less than \SI{100}{\micro\second} [zone B in Fig.~\ref{fig:3}(a)]. Additionally, for the 460~nm pore, a much higher rate of patch heterogeneous nucleation is observed, owing to the lower  cross-pore temperature difference $\Delta T_\mathrm{p}$ corresponding to a given wall temperature $T_\mathrm{w}$. With increasing pore diameter, the specific Joule heat density within the pore volume decreases as the pore volume offers less electrical resistance than the access region~\cite{Gadaleta2014}. This decreases the temperature gradient  from the pore center to the pore walls~\cite{paul2020single}, resulting in a lower $\Delta T_\mathrm{p}$.

Owing to  uncertainty in the heterogeneous nucleation temperature~\cite{Witharana2012}, the waiting times between bubble nucleations are stochastic, leading to nonperiodic bubble signals in the nucleate boiling regime, as shown in Fig.~\ref{fig:1}(c). Consequently, in zone B, a wide or nonspecific frequency band (0--5~MHz) is noticeable in the current and pressure spectrograms [Fig.~\ref{fig:3}(b)]. Each blockage signal in this zone indicates a separate bubble event  comprising bubble nucleation, inertial and evaporation-induced growth to several micrometers, and eventual collapse back to the liquid phase owing to the lack of  sufficient  sensible heat supply. The baseline current is recovered after each bubble collapse. Owing to the abundant  heterogeneous nucleation inside the 460~nm pore, the probability of eventual coalescence of patch nuclei toward  formation of a film-like torus bubble on the pore surface is increased compared with the 420~nm pore. As a result, zone B is short-lived, and is followed by zone C-i at 6.6~V.

The film-to-nucleate boiling transition is activated at higher bias voltages and is best captured at 7~V. As the voltage is increased from 6.6~V to 7~V, a steady and pseudostable oscillating torus bubble is never seen, as is evident from the continuously varying baseline current [Fig.~\ref{fig:3}(d)]. Also, the spectrograms reveal intermittent narrow frequency bands (C-ii) separated by unstable torus oscillations (C-i) and nucleate boiling (B) [Fig.~\ref{fig:3}(e)]. This signifies that the torus bubble is only temporarily stable, and boiling switches chaotically~\cite{shoji2004studies} between nucleate boiling and film boiling multiple times. This effect is similar to the intermittent film boiling seen during transition pool boiling before the Leidenfrost point. From the first work by Nukiyama onward, much attention has been paid to  intermittent film boiling, but an overall model has yet to be established~\cite{Dhir2013}. Actually, for many years, the sudden dip in  the boiling curve from the critical heat flux  (CHF) until  the Leidenfrost point  was generally represented by a dashed and broken  line rather than a well-characterized continuous curve~\cite{lienhard1985historical}. In this paper, we show that by precisely controlling the bias voltage and capturing fast current transitions with a high-bandwidth oscilloscope, this region can now be characterized in minute detail.

\begin{figure}
\centering
\includegraphics[width=1\columnwidth,keepaspectratio,angle=0]{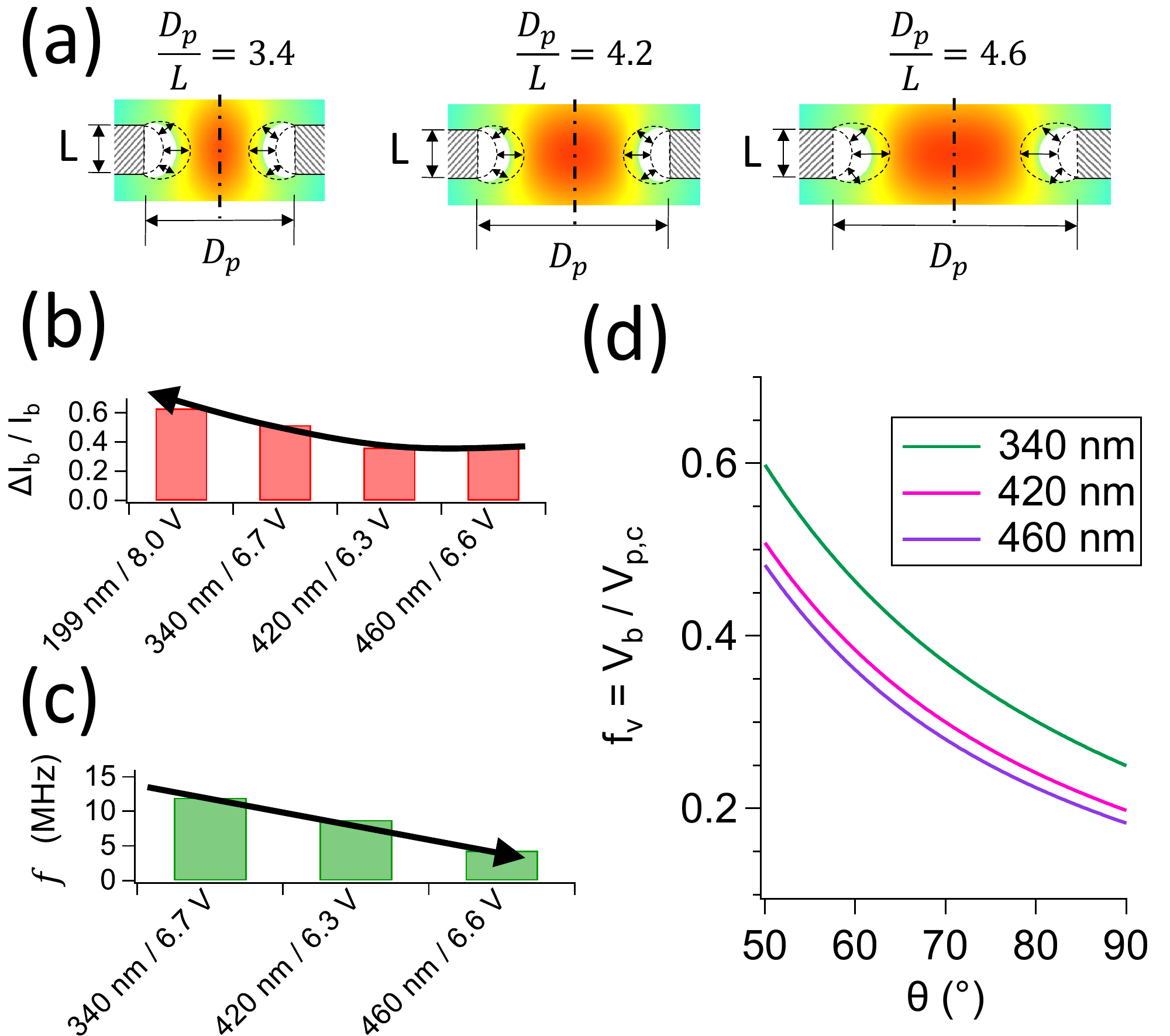}
\caption{\label{fig:35} (a) Schematic showing the torus bubble within nanopore for varying pore diameters [340 nm, 420 nm, 460 nm]. (b) Current baseline dip ratio ($\Delta I_\text{b}/I_\text{b}$) and (c) first harmonic frequency of stable or pseudo-stable torus bubble oscillations as seen in experiments for the three pore diameters. (d) Volume exclusion ratio $f_\text{v}=V_\text{b}/V_\text{p,c}$ variation with contact angle for the three pore diameters. $V_\text{b}$ is the volume of the torus bubble and $V_\text{p,c}$ is the cylindrical pore volume as shown in Fig.~\ref{fig:4}a.}
\end{figure}

A further increase in the bias voltage to 8~V reveals the strange phenomenon of nanopore boiling reverting to completely nucleate boiling. Figures~\ref{fig:3}(g) and~\ref{fig:3}(h) reveal a recovery in the baseline current   between discrete blockage signals, and the spectrogram in Fig.~\ref{fig:3}(i) reveals wideband oscillation power at 0--5~MHz range akin to the nucleate boiling regime (A) in Figs.~\ref{fig:3}(b) and~\ref{fig:3}(e). The reverse transition (film-to-nucleate) with increasing bias voltage is because of a higher Joule heat generation rate, which destabilizes the torus bubble, causing it to collapse. It should be noted that thermal/vapor bubbles are always in unstable equilibrium~\cite{Prosperetti2017}, and oscillations are stable only for small-amplitude driving forces~\cite{hao1999dynamics}.

As the total Joule heat generation within the pore also scales with the pore diameter, it can be expected that  at a given voltage, the torus bubble will be more stable for smaller pore diameters. This can be observed by comparing the  spectrograms for the 460~nm pore at 7~V [Fig.~\ref{fig:3}(e)] with those for the 340~nm pore at 7~V [Fig.~S14(c)~\cite{supp}]. While the 460~nm pore exhibits intermittent boiling at this voltage, the 340~nm pore exhibits stable film boiling at $\sim$15~MHz for more than 2~ms. Also, we can compare Fig.~S11(b)~\cite{supp}, which shows the spectrogram for an expanded 250~nm pore under 7~V. Steady frequency bands in the 10~MHz range are observed. Because the experiments were repeated multiple times, the pore expanded from its initial size of 250~nm to the 300--400~nm range owing to erosion caused by nanobubble vibrations.

In the Supplemental Material~\cite{supp}, we examine the long-term stability of the torus bubble inside the 460~nm pore at 6.3~V and 6.5~V (see Figs.~S15 and~S16, respectively). We find that the 6.3~V torus bubble underwent highly nonlinear oscillations, with many subharmonic and superharmonic frequency bands. In the case of 6.5~V, the torus bubble oscillated weakly nonlinearly for $\sim$\SI{400}{\micro\second}, before returning to nonspecific frequency bands indicative of its instability. Although we do not exactly understand the nonlinear behavior of these bubbles, we can still safely infer that for the 460~nm pore, the torus bubble is either pseudostable or unstable. Comparing the baseline current dip ratio ($\Delta I_\text{b}/I_\text{b}$) during stable film boiling (Fig.~\ref{fig:35}b), we find that the 340 nm pore has a larger ratio than the 420 nm pore, while the 460 nm pores have nearly the same ratio as the 420 nm pore. This can be explained by comparing the volume exclusion ratios, $f_\text{v}$ of the torus bubble inside the three pore diameters. As the dip in baseline current is proportional to the volume exclusion effect posed by the torus bubble, we can expect that $f_\text{v}\propto\Delta I_\text{b}/I_\text{b}$. As shown in Fig.~\ref{fig:35}d, $f_\text{v}(340 \text{nm})>f_\text{v}(420 \text{nm})\approx f_\text{v}(460 \text{nm})$. This is responsible for the current dip ratio trend shown in Fig.~\ref{fig:35}b. Also, the fundamental frequency of the bubble is observed to increase with decreasing pore diameter (Fig.~\ref{fig:35}c). As the oscillating bubble takes a torus shape, its volume ($V_\mathrm{b}$) will be directly proportional to the pore diameter which will cause the frequency, $f$ to vary inversely with diameter as frequency scales inversely to volume according to~\cite{dockar2020forced}.
In Sec.~\ref{secIVB}, we discuss the underlying mechanism of torus bubble instability under a confined heating and pinning effect using a theoretical model. The effects of pore diameter and bias voltage on stability are also elucidated.


\section{Discussion}\label{secIV}
\subsection{Heat generation during nanopore boiling}\label{secIVA}

\begin{table*}[!t]
\caption{\label{tab2}Variation of Joule heat generation during boiling regimes.}
\begin{ruledtabular}
\begin{tabular}{lcdddddddd}
 &  & \multicolumn{2}{c}{Zone A} & \multicolumn{2}{c}{Zone B} & \multicolumn{2}{c}{Zone C-i} & \multicolumn{2}{c}{Zone C-ii}\\
\cline{3-4}\cline{5-6}\cline{3-4}\cline{7-8}\cline{9-10}
$D_\mathrm{p}$ (nm) & $V_\mathrm{app}$ (V)  &  \multicolumn{1}{c}{$t_\mathrm{z}$~(\SI{}{\micro\second})} & \multicolumn{1}{c}{$H$~(\SI{}{\micro\watt})}& \multicolumn{1}{c}{$t_\mathrm{z}$~(\SI{}{\micro\second})}& \multicolumn{1}{c}{$H$~(\SI{}{\micro\watt})}&\multicolumn{1}{c}{$t_\mathrm{z}$~(\SI{}{\micro\second})}& \multicolumn{1}{c}{$H$~(\SI{}{\micro\watt})}&\multicolumn{1}{c}{$t_\mathrm{z}$~(\SI{}{\micro\second})}&\multicolumn{1}{c}{$H$~(\SI{}{\micro\watt})}\\
\hline
420 & 6.3 &261 & 758 & 239 & 710 & 0 & 0 & 0 & 0\\
& 6.4 & 205 & 772 & 116 & 669 & 65 & 484 & 114 & 517\\
460 & 6.6 &263 & 1187 & 50 & 900 & 627 & 725 & 525 & 777\\
& 7.0 &44 & 1389 & 614 & 984 & 510 & 822 & 297 & 789\\
\end{tabular}
\end{ruledtabular}
\end{table*}
In the previous section, nucleate boiling, transition boiling and film boiling were categorized and analyzed minutely within each voltage pulse by deciphering changes in baseline current and oscillation frequency. As the bubble volume, pore volume, and heating volume are all comparable, the presence of a bubble inside the pore severely restricts Joule heat generation. In this section, we report the effect of the nanopore boiling regime on the heat generation rate. This effect is similar to pool boiling, where the transition to film boiling reduces the heat flux at the solid surface by cutting off liquid contact from the solid wall. However, the critical advantage of studying nanopore boiling is that it allows us to focus on a single nucleation site, which is not possible in the case of pool boiling, where heat transfer properties are spatially averaged over multiple nucleation spots. Another advantage of nanopore boiling is that it allows us to investigate the transient effect of boiling transitions, thus highlighting the nonequilibrium states lying between nucleate boiling and stable film boiling. This property is specifically useful in understanding intermittent film boiling during transition boiling, an area of profound scientific interest since the work of Nukiyama.

Table~\ref{tab2} shows the average Joule heating rate $H=V_\mathrm{app} I$ in different boiling zones for a single voltage pulse of duration $t_\mathrm{p}$. $t_\mathrm{z}$ denotes the total duration of each zone during the pulse. 
For the 420~nm and 460~nm pores, $t_\mathrm{p}$ is taken as \SI{500}{\micro\second} and \SI{1468}{\micro\second}, respectively. The 420~nm and 460~nm pore results were captured using the passive and active probes, respectively.

Some clear trends can be seen in Table~\ref{tab2}. For example, the duration of the initial heating zone (A)  becomes shorter with higher voltage, signifying a more rapid temperature rise to nucleation conditions. Compared with the regime in zone A, $H$ decreases during the nucleate boiling  (B) and film boiling (C) regimes, with the latter exhibiting a greater decrease. For the 420~nm pore, during nucleate boiling (B), $H$ is higher at 6.3~V than at 6.4~V. This is because the periodicity of bubble nucleation increases, i.e., the waiting times decrease, with increasing bias voltage, leading to more ionic current blockages. In the case of the 460~nm pore, we find that the duration of the nucleate boiling zone  (B) is significantly longer at 7~V than at 6.6~V, because nucleate boiling regimes (B) appear repeatedly, suggesting the start of transition boiling. For the same reason, the duration of stable film boiling (C-ii) becomes shorter with increasing voltage.  $H$ increases by 17\% in zone A, but by only 1.5\% in zone C-ii. This is due to the volume exclusion effect imposed by the torus bubble.

\subsection{Nano-torus bubble at quasi-equilibrium}\label{secIVB}
\begin{figure}
\includegraphics[width=\columnwidth,keepaspectratio,angle=0]{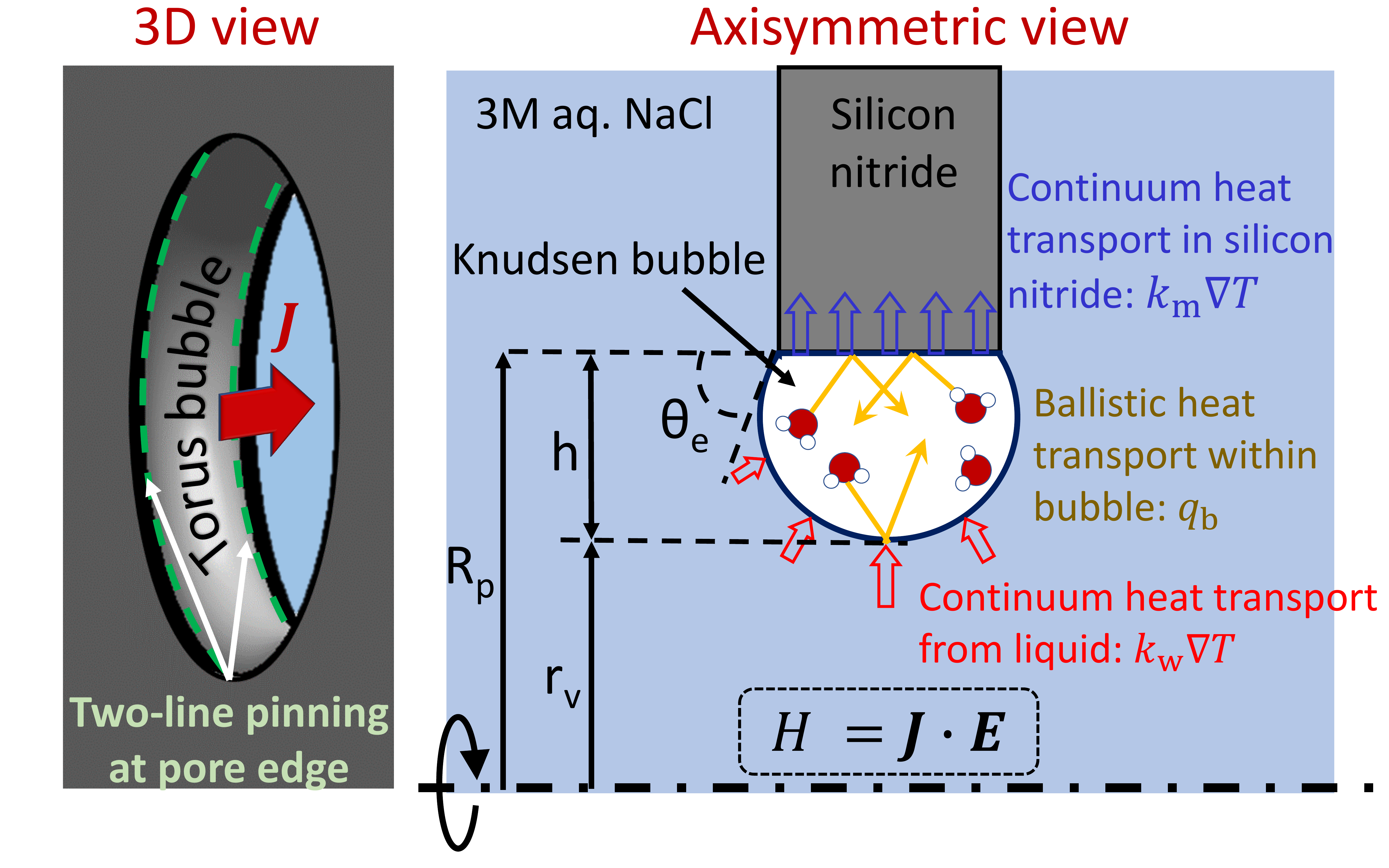}
\caption{\label{fig:41}Schematic explanation of the heat transfer mechanism through the torus bubble sandwiched between the Joule heated liquid and the silicon nitride membrane which acts like a heat sink. The ion flux $\mathbf{J}$ through the unrestricted pore volume generates Joule heat $H$, part of which is transferred through the torus bubble to the silicon nitride walls. Continuum heat transport takes place in the liquid and solid while ballistic heat transport is takes place through the bubble where the mean free path of vapor molecules is in the same order as the bubble height, $h$.}
\end{figure}

\begin{figure*}[!t]
\includegraphics[width=0.8\textwidth,keepaspectratio,angle=0]{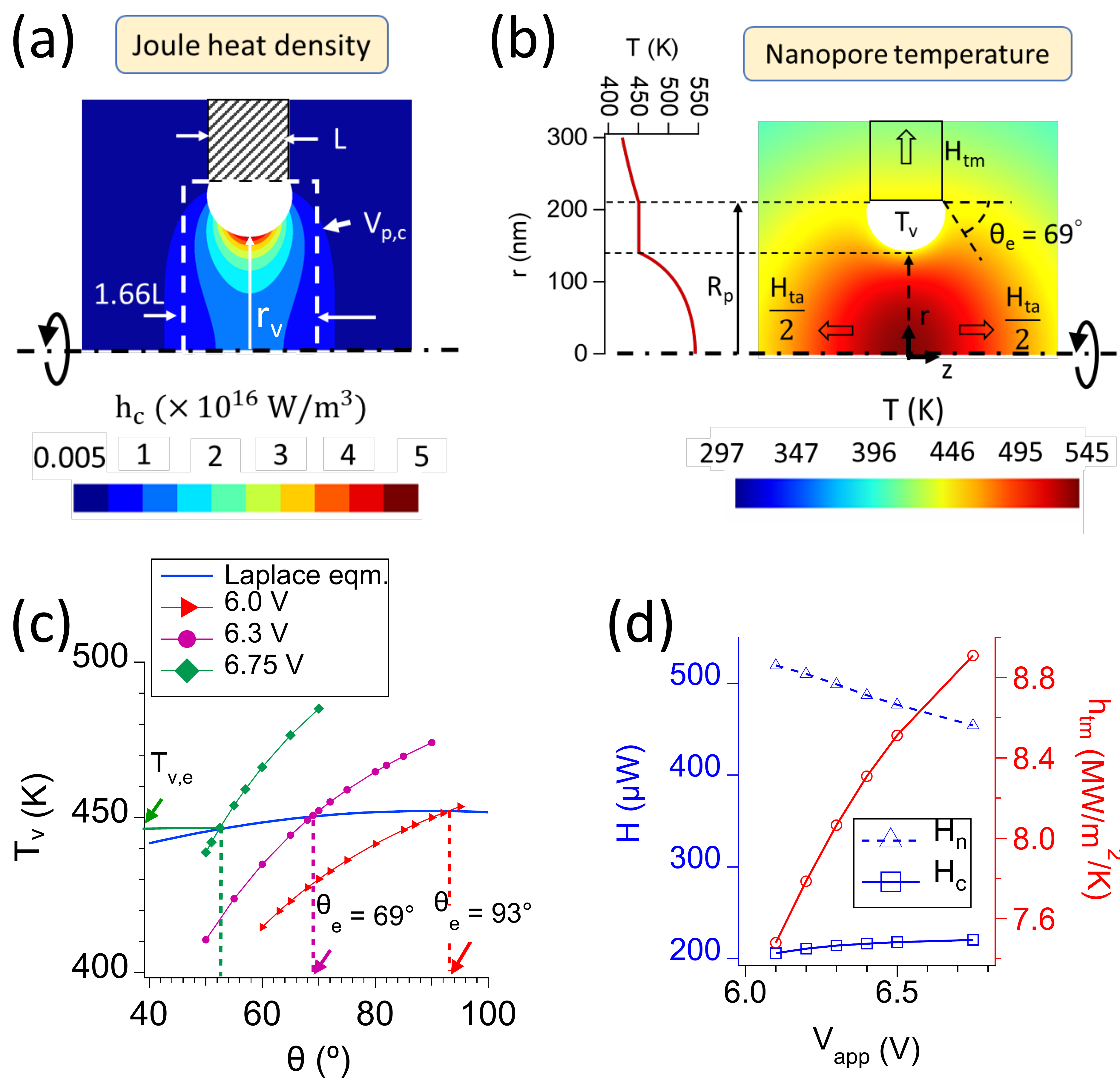}
\caption{\label{fig:4} (a) Steady-state Joule heat density and (b) contour plot of the steady-state temperature distribution for a 420~nm pore under 6.3~V  with a $\theta=69$\degree\ torus bubble sitting on top of  $L=100$~nm thick silicon nitride membrane walls. The temperature distribution along the bubble midline $r_\mathrm{v}$ is shown on the left of (b). (c) Variation of bubble temperature with contact angle under thermal steady-state conditions (lines with markers). The solid blue line shows the variation of bubble temperature under mechanical (Laplace) equilibrium conditions. Solution of the thermal and mechanical equilibrium conditions give the overall equilibrium bubble contact angle $\theta_\mathrm{e}$. (d) Variation of the net Joule heat generation $H_\mathrm{n}$, the Joule heat generation $H_\mathrm{c}$ inside $V_\mathrm{p,c}$, and the heat transfer coefficients of the membrane $h_\mathrm{tm}$ when a torus bubble with equilibrium contact angle $\theta_\mathrm{e}$ occupies the pore volume.}
\end{figure*}

In Sec.~\ref{secIIIA}, we showed that in zone C-ii, steady-state current oscillations are seen with respect to a stable yet reduced baseline. The reduction in baseline current was attributed to the formation of a pseudostable torus bubble with  oscillation amplitude proportional to the volumetric oscillation amplitude of the bubble. In this subsection, we model the mean torus bubble size inside the 420~nm pore by matching the experimental steady-state current in zone C-ii with steady-state Joule heating simulations. As the bubble exists in a pinned state, decreasing the contact angle causes an increase in bubble volume.

When a torus bubble is introduced om the cylindrical silicon nitride pore surface as shown in Fig.~\ref{fig:41} and Fig.~\ref{fig:4}a, the Joule heat density in the liquid increases, while the total Joule heat production decreases, because of the volume exclusion effect. The torus bubble shape is calculated [Appendix~\ref{appA}] assuming that it has a constant mean surface curvature (CMC) and it remains pinned on the two pore edges as shown in Fig.~\ref{fig:41}. The vapor molecules inside the torus bubble transports heat collides between the liquid/vapor and solid/vapor interfaces (axisymmetric view in Fig.~\ref{fig:41}), transporting heat ballistically with no or minor inter-molecular collisions. The ballistic heat flux for a given temperature drop is much more powerful and faster than continuum heat transport in the liquid [Eqs.~\eqref{eq:ap85} in Appendix~\ref{appB}]. Hence, the torus bubble offers no thermal resistance to heat transfer between the liquid/vapor and solid/vapor interfaces. We solve the energy conservation equations [Eqs.~\eqref{eq:ap4} in Appendix~\ref{appB}] in the electrolyte and the silicon nitride membrane, taking into account the Joule heat source term $H=\mathbf{J}\cdot\mathbf{E}\approx\sigma \vert \mathbf{E} \vert^2$, where $\sigma$, $\mathbf{J}$ and $\mathbf{E}$ are the electrical conductivity, ion flux density and electric field, respectively. When the heat generation inside the pore and the heat diffusion through the membrane are balanced [Eqs.~\eqref{eq:ap8} and Fig.~S7(b)~\cite{supp}], the steady-state nanopore and bubble temperatures are obtained [Fig.~\ref{fig:4}(b)]. In this configuration, the bubble has a vapor temperature  $T_\mathrm{v}=$450.2~K. The net Joule heat production, $H_\mathrm{n}$ is the sum total of the heat generation in the cylindrical pore region around the bubble [$H_\mathrm{c}$ inside $V_\mathrm{p,c}$, shown in Fig.~\ref{fig:4}(a)] and the heat generation in the access region away from the bubble ($H_\mathrm{a}$), i.e., $H_\mathrm{n}=H_\mathrm{c}+H_\mathrm{a}$. For $\theta_\mathrm{e}=69\degree$, as shown in Fig.~\ref{fig:4}(b), $H_\mathrm{n}=499$~\SI{}{\micro\watt}, $H_\mathrm{c}=214$~\SI{}{\micro\watt}, and $H_\mathrm{a}=283$~\SI{}{\micro\watt}. $H_\mathrm{c}$ is balanced by the diffusive heat flux through the liquid surrounding the pore  ($H_\mathrm{ta}=52$~\SI{}{\micro\watt}) and the diffusive heat flux through the silicon nitride membrane ($H_\mathrm{tm}=162$~\SI{}{\micro\watt}). It should be noted that the majority of the Joule heat ($\upsilon=H_\mathrm{tm}/H_\mathrm{c}=0.76$ or 76\%) produced within the cylindrical pore region is consumed by the bubble and transferred to the silicon nitride surface on top of which it sits. Also, $H_\mathrm{tm}=h_\mathrm{tm} 2\pi R_\mathrm{p}L (T_\mathrm{v}-T_0)$, which can be solved to obtain the heat transfer coefficient of the membrane at $V_\mathrm{app}=6.3$~V, namely,  $h_\mathrm{tm}=8.06$~MW\,m$^{-2}$\,K$^{-1}$ [Fig.~\ref{fig:4}(d)].

Figure~\ref{fig:4}(c) shows the variation of torus bubble temperature with contact angle according to mechanical equilibrium (blue line) and thermal steady-state conditions (lines with markers) for different voltages. The vapor temperature for mechanical equilibrium has been obtained according to Eq.~\eqref{eq:ap9} in Appendix~\ref{appC}, assuming  saturated vapor pressure inside the bubble, which is balanced by the Laplace pressure. In the thermal steady state, there is no heat accumulation in the bubble, and also there exists no temperature drop across the interface to cause a net evaporation flux [Eqs.~\eqref{eq:ap8}]. We can see that for a given voltage, there is only one solution for the contact angle at which both mechanical equilibrium and steady-state conditions are simultaneously satisfied, namely, $\theta_\mathrm{e}=69\degree$ when $V_\mathrm{app}=6.3$~V. This steady-state size and temperature of the bubble can also be termed  a quasi-equilibrium state~\cite{maheshwari2018dynamics}. It should be noted here that the growth--collapse cycle of nucleate bubble (either homogeneous or heterogeneous) is a highly transient and out-of equilibrium process~\cite{paul2021analysis}, and the torus film bubble can reach a quasi-equilibrium state that is experimentally observed to be stable for longer than \SI{100}{\micro\second}. This is due to the high volume of the torus bubble, which restricts Joule heat production within the pore and arrests the transient liquid temperature rise, enabling a thermal steady state to exist. It is interesting to note that at $\theta_\mathrm{e}=69\degree$, the nanopore current from the simulations [Fig.~S6(a)~\cite{supp}] and the mean oscillation current observed experimentally [$I_\mathrm{b,a,e,Q}$ in Fig.~S17(c)~\cite{supp}] are in agreement with each other. This justifies the attribution of the baseline current dip to the amount of volume occlusion in the pore volume provided by the torus bubble.

It should also be noted that with increasing voltage, the equilibrium temperature of the bubble $T_\mathrm{v,e}$ remains almost constant, while the pinned bubble grows (i.e., $\theta_\mathrm{e}$ decreases) significantly. Because of the negative curvature of the pore surface and the pinning effect, the curvature of the torus bubble is less sensitive to decreasing contact angle [pinned expansion, see Fig.~S3(c)~\cite{supp}], as a result of which the vapor temperature as given by the Laplace equation remains quasi-uniform. Also, the heat transfer capacity of the silicon nitride membrane is limited, which forces the bubble to grow to the limit $H_\mathrm{c}$ such that a thermal steady state can be established. This can be explicitly observed in Fig.~\ref{fig:4}(d), where for $\theta = \theta_\mathrm{e}$, the net Joule heat generation $H_\mathrm{n}$ decreases while $H_\mathrm{c}$ increases slightly with increasing bias voltage. Qualitatively, this finding supports our experimental results presented in Table~\ref{tab2}, where increasing the voltage from 6.6~V to 7~V caused $H$ to rise by only  1.5\% in zone C-ii. This analysis shows that, similar to vapor films in pool boiling, which impede the solid-to-liquid heat flux, nanopore torus bubbles  also exert a limiting effect, but on  Joule heat generation in the bulk liquid, through volume exclusion. As  Joule heating is restricted by the torus bubble, the pore center temperature is unable to rise to 600~K, thereby preventing homogeneous nucleation.

\subsection{Stability of Nano-torus bubble}\label{secIVC}

\begin{figure*}[!t]
\includegraphics[width=0.8\textwidth,keepaspectratio,angle=0]{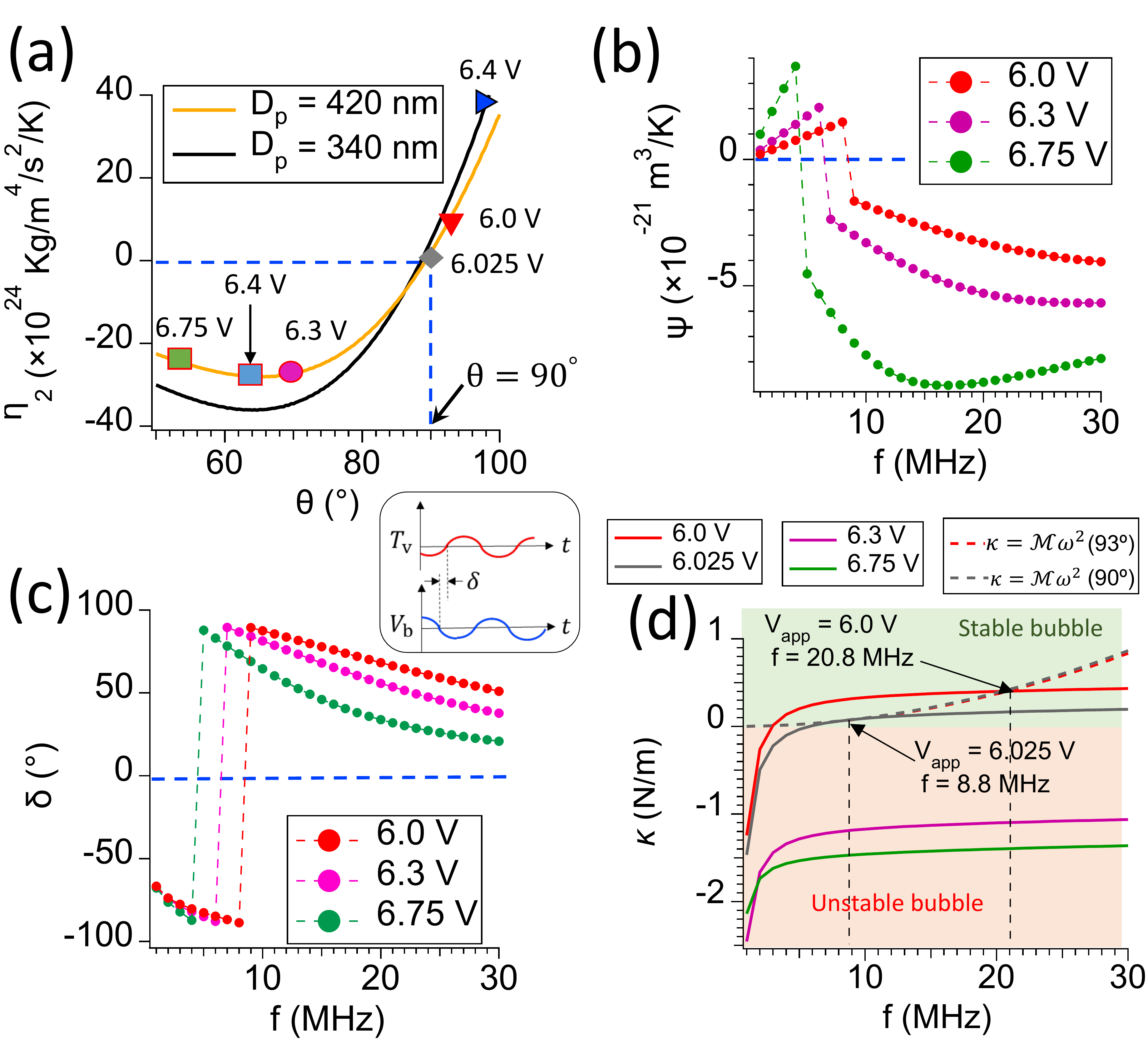}
\caption{\label{fig:5} Variation of torus bubble stability parameters with oscillation frequency, $f$ under different bias voltages $V_\mathrm{app}$: (a) curvature coefficient $\eta_2$; (b) volume expansion coefficient $\psi$; (c) phase difference $\delta$ between bubble temperature and volume; (d) mechanical stiffness $\kappa$.}
\end{figure*}

Using nanopore Joule heating, we reduced the boiling scale to the 100~nm range, thus entering the domain of single-bubble dynamics. At this scale, buoyancy become insignificant, while surface tension dominates, as characterized by the very low Bond number $Bo=\rho_\mathrm{w}gR^2/\gamma = \mathcal{O}(10^{-9})$. Also, hydrodynamic instabilities can be ruled out at this scale, since the critical wavelengths for surface waves according to Rayleigh--Taylor ($\lambda_\mathrm{c}= 14$~cm) and Kelvin--Helmholtz  ($\lambda_\mathrm{c}= 2.3$~mm) instabilities~\cite{carey2020liquid}are four orders of magnitude larger than the dimensions of the bubble. For Kelvin--Helmholtz instability, we have assumed zero vapor velocity ($u_\mathrm{v}= 0$) because of the pinning effect, while we have assumed the velocity of the liquid surrounding the bubble  to be equal to the typical electro-osmotic flow velocity inside the nanopore,  $u_\mathrm{l}= 5$~m/s~\cite{wang2020joule}.

We judge the stability of thermal bubble equilibrium from first principles, by perturbing the bubble slightly and checking whether there is a positive restoring force~\cite{hao1999dynamics, prosperetti2015speed}. Hence, when a thermal bubble initially in quasi-equilibrium is radially perturbed ($\Delta R$), the restoring force $F$ can be written as
\begin{equation}
F=S_\mathrm{b}\Delta(P_\mathrm{v} - P_\mathrm{w} - K\gamma ) = -\kappa \Delta R.
\label{eq:two}
\end{equation}
Here, $K$ and $S_\mathrm{b}$ are the curvature and liquid/vapor interfacial surface area of the bubble, respectively  (Fig.~S3~\cite{supp}). $\kappa$ is the mechanical stiffness, which must satisfy $\kappa>0$ for stable equilibrium. $P_\mathrm{v}$ and $\gamma$ are the vapor pressure and surface tension, respectively, of water at saturation temperature. $P_\mathrm{w}$ is the atmospheric pressure. 
It should be noted that submicrometer-scale thermal spherical bubbles are regarded to be in unstable equilibrium~\cite{Prosperetti2017}. For a spherical vapor bubble with radius $R_\mathrm{e}$ and temperature $T_\mathrm{v}$ in mechanical and thermal equilibrium in an infinite reservoir, the linear oscillation model, Eq.~\eqref{eq:two}, can be written as~\cite{prosperetti2015speed}
\begin{equation}
\rho_\mathrm{w}R_\mathrm{e}{\omega}^2= \frac{h_\mathrm{fg}\rho_\mathrm{v}}{\rho_\mathrm{w}c_\mathrm{w}T_\mathrm{v}}\sqrt{\frac{\omega}{2D_\mathrm{th}}} - \frac{2\gamma}{{R_\mathrm{e}}^2},
\label{eq:four}
\end{equation}
with $\kappa=\mathcal{M}{\omega}^2$, where $\mathcal{M}=4\pi\rho_\mathrm{w}R_\mathrm{e}^3$ is the reduced mass of the spherical bubble~\cite{Vincent2017}. There exists a real solution for $\omega$ only when $R_\mathrm{e}\geq \SI{14}{\micro\meter}$. This is due to the inverse relation between curvature and bubble radius, $K=2/R$, which causes the surface tension force to dominate for smaller bubbles, leading to $\kappa<0$ and preventing stable equilibrium. This analysis, however, begs the question: how does a $\mathcal{O}(100~\mathrm{nm})$ torus bubble attain stability within the nanopore, making stable film boiling possible in the first place.

Next, we judge the thermal stability of the torus bubble equilibrium. Assuming that  thermal equilibrium is maintained at the liquid--vapor and solid--vapor interfaces, ($T_\mathrm{v}=T_\mathrm{w}=T_\mathrm{m}$) and that the vapor temperature  varies along the saturation line ($\Delta P_\mathrm{v} / \Delta T_\mathrm{v} = h_\mathrm{fg}\rho_\mathrm{v}/T_\mathrm{v} $)~\cite{hao1999dynamics}, Eq.~\eqref{eq:two} can also be written as
\begin{align}
\kappa &= -\frac{S_\mathrm{b}^2}{\psi}\!\left(\frac{h_\mathrm{fg}\rho_\mathrm{v}}{T_\mathrm{v}}\cos\delta-\psi\gamma\frac{dK}{dV_\mathrm{b}}\right)\nonumber \\[3pt]
&= {S_\mathrm{b}^2}\!\left(\eta_2 - \frac{h_\mathrm{fg}\rho_\mathrm{v}}{T_\mathrm{v}}\frac{\cos\delta}{\psi}\right)=\mathcal{M(\theta_\text{e})}{\omega}^2.
\label{eq:three}
\end{align}
$\mathcal{M(\theta_\text{e})}$ is the reduced mass of the torus bubble having quasi-equilibrium contact angle, $\theta_\text{e}$. A detailed derivation is provided in Appendix~\ref{appD} [Eqs.~\eqref{eq:ap10}--\eqref{eq:ap171}]. $h_\mathrm{fg}$ and $\rho_\mathrm{v}$ are the latent heat and vapor density, respectively, assuming saturation conditions. For linear oscillations of the bubble volume $V_\mathrm{b}$ and temperature $T_\mathrm{v}$ given by $V_\mathrm{b} = V_\mathrm{b,e} + V_\mathrm{b,a}\exp(j\omega t)$ and
$T_\mathrm{v} = T_\mathrm{v,e} + T_\mathrm{v,a}\exp[j(\omega t + \delta)]$, with $j=\sqrt{-1}$, $\psi=V_\mathrm{b,a}/T_\mathrm{v,a}$ denotes the volume expansion coefficient (i.e., the ratio of volume and temperature amplitudes) and $\delta$ ($-90\degree\leq\delta\leq90\degree$) denotes the phase difference between them. $\eta_2$ is the curvature coefficient capturing the variation of bubble curvature with contact angle for a given pore size as shown in Fig.~S4~\cite{supp}.

Thus, for stable equilibrium, the mechanical stiffness of the bubble must satisfy $\kappa>0$, which necessitates that there be a positive restoring force responsible for returning the bubble back to equilibrium conditions after it has been perturbed. As the bubble expands, the Joule heat density inside the pore increases [$d\hat{h}/d\theta<0$ in Fig.~S6(c)~\cite{supp}]. This rise in heat generation is balanced by the internal energy rise of the system, the latent heat of evaporation into the bubble, the $P$--$V$ work done by the bubble, and the rise in heat flux from the bubble to the silicon nitride membrane [Eq.~\eqref{eq:ap12} in Appendix~\ref{appD}]. Thus, by perturbing the energy balance of the bubble, $\psi$ and $\delta$ are obtained. Excess heat generation leads to $\psi>0$. On the other hand, if the bubble expansion consumes more heat than is being generated, then $\psi<0$.

Figure~\ref{fig:5} summarizes the four stability parameters $\eta_2$, $\psi$, $\delta$, and $\kappa$ for the equilibrium torus bubble inside the 420~nm pore. Owing to the pinning effect, the curvature perturbation $\eta_2\geq0$ for $\theta_\mathrm{e}\geq$ 90\degree, which causes $\kappa$ to increase [Eq.~\eqref{eq:three}], providing a positive restoring force following volume perturbation. In this $\theta$ range, at the high-frequency limit where $\psi<0$, temperature perturbations will also tend to increase $\kappa$, thus lending an extra stabilization effect. Owing to the dual effect of curvature stability ($\eta_2>0$) and thermal stability ($\psi<0$), nano-torus film bubbles inside a nanopore can exist in stable equilibrium. This behavior is, however, not possible for spherical bubbles at the same scale.

However, when $\psi>0$ [Fig.~\ref{fig:5}(b)] and $\vert\delta\vert<90\degree$ in the low-frequency range [Fig.~\ref{fig:5}(c)], the thermal perturbation 
\[
-\frac{h_\mathrm{fg}\rho_\mathrm{v}}{T_\mathrm{v}}\frac{\cos\delta}{\psi}
\]
 becomes negative and can overpower the curvature stability, triggering destabilization. This can be seen for the $\kappa$ trace [Fig.~\ref{fig:5}(d)] at  6.0~V, which enters the stable equilibrium zone ($\kappa>0$) only beyond a critical frequency (3~MHz). In the stable equilibrium zone, the solution for $\kappa$ according to Eq.~\eqref{eq:three} (solid line) and $\kappa = \mathcal{M}{\omega}^2$ [dashed line in Fig.~\ref{fig:5}(d)] gives a resonant frequency  $ f =20.8$~MHz for the 93\degree~torus bubble at 6.0~V and $ f =8.8$~MHz for the 90\degree~torus bubble at 6.025~V. Here, $\omega=2\pi f $ is the angular frequency and $\mathcal{M}=2.23\times10^{-17}$~kg is the reduced mass for the 93\degree~torus bubble inside the 420~nm pore. The method for calculating the reduced mass numerically is described in  Appendix~\ref{appE} [Eq.~\eqref{eq:ap20}]. By comparison, in experiments, we observed a stable oscillation frequency of 8.9~MHz for the 420~nm pore at 6.3~V. Compared with the 93\degree~torus, the curvature coefficient $\eta_2$ is smaller for the 90\degree~torus, leading to a smaller stiffness and resonant frequency. Again, for $\theta<90\degree$, $\eta_2$ decreases, leading to bubble instability. Hence, we can say that $ f =8.8$~MHz is the minimum frequency for stable torus oscillations. It should also be noted that at $\theta_\mathrm{e}=90\degree$, the curves of $\kappa = \mathcal{M}{\omega}^2$ and Eq.~\eqref{eq:three} are tangent to each other. Also, at this limit where the modeled and experimental resonant frequencies match, the error of $\sim$0.3~V  is acceptable given the approximations in our model. We should also cite the analytical model by Dockar \emph{et al.}~\cite{dockar2020forced}, which was formulated for adiabatic bubble oscillations on a flat surface. According to that model, the natural frequency of a 90\degree~patch heterogeneous bubble pinned on a 100~nm flat surface is in the region of 450~MHz. The one order smaller frequency for the torus bubble observed here is due to (i) a larger volume compared with a patch bubble, which leads to a larger reduced mass $\mathcal{M}$, and (ii) lower stiffness and smaller volume expansion coefficient magnitude $\vert\psi\vert$ owing to Joule heat density amplification with bubble enlargement.

However, with increasing bias voltage, the equilibrium bubble volume increases (i.e., $\theta_\mathrm{e}$ decreases), which causes the bubble to enter  the $\eta_2<0$ zone, where the surface tension force tends to reduce $\kappa$. This can be observed in Fig.~\ref{fig:5}(a), where for 6.0~V, $\eta_2>0$, but for 6.3~V and 6.75~V, $\eta_2<0$. In this range, even for $\psi<0$ and $\delta\xrightarrow{}0\degree$,  a positive $\kappa$ is not obtained, and thus  the torus bubble equilibrium is unstable. This trend is supported by our experiments, where with increasing voltage, the stable torus bubble zone (C-ii) decreases in extent and eventually disappears, paving the way toward transition back to nucleate boiling.

This theory also explains the higher stability of torus bubble films with decreasing pore size. As shown in Fig.~\ref{fig:5}(a), for the 340~nm pore, a bias voltage of 6.4~V results in an equilibrium torus bubble contact angle in the $\theta_\mathrm{e}>90\degree$ range, where curvature perturbation enhances bubble stability ($\eta_2>0$). As the heat transfer capacity of the silicon nitride membrane ($h_\text{tm}$) is limited, the bubble volume would increase to limit $\text{H}_\text{c}$ rise with increasing voltage, so that thermal steady-state is achieved. 
Therefore, smaller pore diameters, which have a higher bubble volume exclusion ratio [$f_\mathrm{v}$ in Fig.~\ref{fig:35}d] for the same $\theta$, allow thermal steady-state to be established at $\theta_\mathrm{e}\xrightarrow[]{}90\degree$ or $\theta_\mathrm{e}>90\degree$. This indicates that stable film boiling will persist to higher voltages as the pore diameter is decreased. Our experiments also support this hypothesis, where stable film boiling exists for 6.7~V and 7~V for the 340~nm pore, whereas for the 420~nm  pore, loss of stable film boiling is triggered at 6.5~V.




It should be noted that this model only checks  stability in the quasi-equilibrium state of the bubble. To avoid the zone of curvature instability ($\theta_\mathrm{e}<90\degree$), the mean bubble size may exist in an out-of-equilibrium state, with $\theta_\mathrm{e}\approx90\degree$, where it can be quasi-stable with nonlinear and large-amplitude oscillations as seen in experimental spectrograms. The linear model described here is not able to capture the complex  out-of-equilibrium behavior of the torus bubble, and a complete multiphase computational fluid dynamics (CFD) treatment may be necessary.

\section{Conclusions}\label{secV}
In summary, we have presented a consolidated picture of boiling inside a cylindrical nanopore, which differs significantly from traditional pool boiling owing to the confined nature of Joule heat generation and bubble dynamics. Measuring the ionic current flow reduction with bubble formation within the nanopore volume and stress waves generated by bubble motion enable investigation of vapor nanobubbles at nanosecond resolution. Two modes of boiling are detected: (i) nucleate boiling inside the nanopore, involving  discrete homogeneous and patch heterogeneous bubble nucleation and (ii) oscillatory film boiling, involving   a torus bubble that restricts Joule heat generation, thus preventing nucleate boiling at the same time. Interestingly, with increasing bias voltage, a reverse transition from film to nucleate boiling is observed. The dynamics of the torus bubble are explained with the help of theoretical models. Owing to the pinning effect, the torus bubble, albeit at the nanoscale, is able to attain a stable equilibrium in the low-bias-voltage range. However, with increasing voltage, the labile vapor bubble expands in a pinned state to limit Joule heat generation and establish quasi-thermal equilibrium. However, the mechanical stiffness of the bubble decreases with increasing volume, and beyond a critical voltage, the stiffness becomes negative, making the equilibrium unstable and ultimately destabilizing the torus bubble. 

\begin{acknowledgments}
This work was supported by the Japan Society for the Promotion of Science (JSPS) KAKENHI Grant Nos. 20H02081 and 20J22422. Part of this work was conducted at the Advanced Characterization Nanotechnology Platform of the University of Tokyo, which was supported by the Nanotechnology Platform of the Ministry of Education, Culture, Sports, Science, and Technology (MEXT), Japan, Grant No. JPMXP09A21UT0089.
\end{acknowledgments}

\appendix
\section{Torus bubble shape calculations}\label{appA}
Assuming that the torus bubble has uniform mean curvature $K$, this can be expressed as~\cite{gray2017modern}
\begin{equation}
K = \frac{b'\sin\nu+b\cos\nu}{(a + b\cos\nu)\sqrt{b^2+b'^2}}- \frac{bb''-2b'^2-b^2}{(b^2+b'^2)\sqrt{b^2+b'^2}},
\label{eq:ap2}
\end{equation}
where $b$ is the radial distance from the first principal center of curvature, which varies with $\nu$. Thus, $b'=db/d\nu$. From symmetry, we know that $b'=0$ when $\nu=180\degree$. Let the value of $b$ at $\nu=180\degree$ be $b_\pi$. Now, for a given value of $K$, we can solve for $b''$, which we can use to incrementally obtain $b$ and $b'$ for decreasing values of $\nu$ from $\nu=180\degree$ to $\nu=0\degree$, thus tracing the constant-mean-curvature (CMC) torus bubble interface. Mathematically,
\begin{equation}
\begin{aligned}
b'(\nu+\Delta \nu) &= b'(\nu) + b''(\nu)\Delta \nu,  \\[3pt]
b(\nu+\Delta \nu) &= b(\nu) + b'(\nu+\Delta \nu) \Delta \nu,
\end{aligned}
\label{eq:ap2point1}
\end{equation}
where $\Delta \nu = -0.0018\degree$. We find that only when $K=(a - 2b_\pi)/[b_\pi(a - b_\pi)]$ is the torus bubble surface  free of perturbations and the surface area is minimized as shown. Now, for different values of $[a,b_\pi]$, $[K,l,\theta]$ are calculated. Here, $l$ is the bubble length on the pore surface and $\theta$ is the contact angle, as shown in Fig.~S3(a)~\cite{supp}. The values of $[a,b_\pi]$ that satisfy the pinning condition $l=L$ are selected, and  geometric properties such as the bubble volume $V_\mathrm{b}$ and surface area $S_\mathrm{b}$ are obtained by numerical integration over the tracer points ($R,Z$) used to construct the surface [Figs.~S3(e) and~S3(f)~\cite{supp}].

\section{Joule heating model}\label{appB}
To obtain the temperature distribution responsible for the bubble behavior, numerical simulations are employed to solve the energy-conservation equations in the liquid and the silicon nitride membrane:
\begin{gather}
\begin{aligned}
\rho_\mathrm{w} c_\mathrm{p,w}\frac{\partial T}{\partial t} &= \frac{1}{r}\frac{\partial}{\partial r}\!\left(k_\mathrm{w} r \frac{\partial T}{\partial r}\right) +\frac{\partial}{\partial z}\!\left(k_\mathrm{w} \frac{\partial T}{\partial z}\right) + \sigma{\vert\mathbf{E}\vert}^2,  \\[12pt]
\rho_\mathrm{m} c_\mathrm{m}\frac{\partial T}{\partial t} &= \frac{1}{r}\frac{\partial}{\partial r}\!\left(k_\mathrm{m} r \frac{\partial T}{\partial r}\right) +\frac{\partial}{\partial z}\!\left(k_\mathrm{m} \frac{\partial T}{\partial z}\right).
\end {aligned}
\label{eq:ap4}\raisetag{36pt}
\end{gather}
Here, $\rho_\mathrm{w}$, $c_\mathrm{p,w}$, and $k_\mathrm{w}$ are the temperature-dependent water density, specific heat, and thermal conductivity, respectively \cite{Levine2016,wagner2002iapws}, and $t$ is time. $\rho_\mathrm{m}$, $c_\mathrm{m}$, and $k_\mathrm{m}$ are the density, specific heat and thermal conductivity of the silicon nitride membrane. $\sigma$ is the electrical conductivity of the electrolyte, which is captured using an empirical relation first established by Levine \emph{et al.}~\cite{Levine2016},
\begin{equation}
\sigma=mT-b-\frac{(T-T_0)^\alpha}{\beta},
\label{eq:ap5}
\end{equation}
where $m=0.391$~S\,m$^{-1}$\,K$^{-1}$, $b=0.391$~S/m, $\alpha=2.65$, $\beta=5.6\times10^4$, and $T_0=293.15$~K. Here, $\alpha$ is the only fitting parameter that is varied to fit the pre-boiling baseline current $I_\mathrm{b}$ of the nanopore.
The electric field inside the liquid is obtained by combining the ion flux balance ($\nabla \cdot \mathbf{J}=d\rho_\mathrm{e}/dt$) and Poisson's equation ($\nabla \cdot \epsilon\mathbf{E} =  \rho_\mathrm{e}/\varepsilon_0$):
\begin{equation}
\nabla \cdot (\sigma \nabla U ) = \frac{d}{dt}[\varepsilon_0 \nabla \cdot (\epsilon U)],
\label{eq:ap6}
\end{equation}
where $\mathbf{J}=\sigma\mathbf{E}$ is the ionic flux, and $U$,  $\rho_\mathrm{e}$, $\varepsilon_0$, and $\epsilon$ are the electric potential, induced charge, dielectric permittivity of free space, and temperature-dependent dielectric constant of water, respectively~\cite{paul2020single, Levine2016}. The boundary conditions for $T$ and $J$ applied on the simulation boundaries are shown in Figs.~S7(a) and~S7(b)~\cite{supp}.

In an axisymmetric reference frame with origin at the center of the nanopore, $r$ represents the radial coordinate and $z$ the axial coordinate. Equations~\eqref{eq:ap4} are solved on a finite-volume mesh with appropriate boundary conditions and numerical discretizations~\cite{Harvie2012} (Sec.~S3 of the Supplemental Material~\cite{supp}). As the nanopore temperature increases, so too does the electrolyte conductivity, allowing greater current flow for the same voltage:
\begin{equation}
\begin{aligned}
I_\mathrm{b} &= 2\pi \int_{0}^{R_\mathrm{p}}{\sigma \mathbf{E}}\cdot \mathbf{\hat{z}}  r \,dr,\\[6pt]
I_\mathrm{b,a} &= 2\pi \int_{0}^{r_\mathrm{v}}{\sigma \mathbf{E}}\cdot \mathbf{\hat{z}}  r \,dr,
\end{aligned}
\label{eq:ap7}
\end{equation}
where $I_\mathrm{b}$ denotes the baseline current development before the onset of bubble nucleation as shown by the red dashed line in Fig.~S17(a)~\cite{supp}. As can be seen, the experimental and simulation current development match reasonably well except for the initial \SI{20}{\micro\sec}, when membrane capacitive charging affects the nanopore current~\cite{paul2020single}. At the point of first homogeneous bubble nucleation, the temperature at the pore center reaches 584~K, which is the range of theoretical homogeneous nucleation temperatures~\cite{Avedisian1985}.

Next, we apply the same model to a nanopore-bubble system where a torus bubble blankets the cylindrical pore surface, as shown in Fig.~\ref{fig:4}(b). We assume that the liquid and silicon nitride are in thermal equilibrium with the bubble temperature, i.e.,
\begin{equation}
\begin{gathered}
T_\mathrm{v} =  T\|_{S_\mathrm{b}} = T\|_{S_\mathrm{p}},\\[6pt]
\int_{S_\mathrm{b}}{k_\mathrm{w} \nabla T \cdot \mathbf{\hat{n}} \,dS}  = \int_{S_\mathrm{p}}{k_\mathrm{m} \nabla T \cdot \mathbf{\hat{n}}\, dS},
\end{gathered}
\label{eq:ap8}
\end{equation}
where $\mathbf{\hat{n}}$ is the normal vector to the surface concerned, $S_\mathrm{b}$ denotes the liquid--vapor interface and $S_\mathrm{p}=2\pi R_\mathrm{p} L$  the vapor-solid interface. Equations~\eqref{eq:ap8} signify that the vapor inside the bubble has a uniform temperature from the liquid/vapor interface to the vapor/solid interface and that the heat flux from the liquid to the bubble is balanced by the heat flux from the bubble to the silicon nitride wall on top of which it sits.

The above assumptions are valid when the vapor inside the bubble is assumed to be a Knudsen gas (i.e., the Knudsen number $\mathrm{Kn}>1$)~\cite{maheshwari2018dynamics, seddon2011knudsen}. According to Craig~\cite{Craig2011},  bubbles smaller than 100~nm typically satisfy the Knudsen gas definition. The Knudsen number is given by $\mathrm{Kn} = \lambda/h$, where $\lambda={k_\mathrm{B}T_\mathrm{v}}/{\sqrt{2}\sigma_\mathrm{A}P_\mathrm{v}}$ is the mean free path of vapor molecules and $h$ is the distance between the liquid/vapor and vapor/solid interfaces as shown in Fig.~\ref{fig:41}. $\sigma_\mathrm{A}$ is the molecular cross-sectional area and $k_\mathrm{B}$ is  Boltzmann's constant. Taking the diameter of a water molecule to be $d_\mathrm{w}=2.7$~\AA\ and $\sigma_\mathrm{A}=\pi{d_\mathrm{w}}^2$, $\lambda=20.6$~nm. For the bubble shown in Fig.~\ref{fig:41}, $h\sim R_\mathrm{p}-r_\mathrm{v}=69$~nm, which leads to a Knudsen number $\mathrm{Kn}= 0.299$, which falls in the transition regime ($0.1<\mathrm{Kn}<10$). On the other hand, the collision length of liquid water molecules in the pore is 0.13~nm, which is much smaller than the pore diameter (340--460~nm). In this situation, the heat transfer in the pore liquid is governed by continuum transport. On the other hand, the heat flux through the bubble can be assumed to be governed by the ballistic heat flux according to kinetic theory ($q_\mathrm{b}$), which is given by~\cite{maheshwari2018dynamics}
\begin{equation}
q_\mathrm{b} = \alpha_\mathrm{c}\rho_\mathrm{w,n}\sqrt{\frac{2k_\mathrm{B}^3}{m}}\,\big(T_\mathrm{lv}^{3/2}-T_\mathrm{sv}^{3/2}\big).
\label{eq:ap85}
\end{equation}
Here, $\alpha_\mathrm{c}$ is the accommodation coefficient, which can be assumed to be 1~\cite{aursand2019comparison}. $m$ is the mass of one water molecule. $\rho_\mathrm{w,n}$ is the liquid molecule number density on the liquid/vapor interface. $T_\mathrm{lv}$ and $T_\mathrm{sv}$ are the temperatures on the liquid/vapor and solid/vapor interfaces. For a Knudsen gas at $T_\mathrm{lv}=450$~K and $T_\mathrm{sv}= 445$~K, $q_\mathrm{b}=\SI{1982.4}{M\watt\per\meter^2}$. On the other hand, the heat flux on the solid/vapor interface, $k_\mathrm{m} \nabla T$, in Fig.~\ref{fig:4}(b) comes out to be \SI{1227}{M\watt\per\meter^2}. These calculations show that a mere 5~K temperature drop between the two interfaces can account for the huge heat flux through the bubble. Physically, this high heat transfer rate is possible when vapor molecules evaporating from the liquid/vapor interface at high temperature travel to and collide with the solid/vapor interface at a lower temperature without interacting with other vapor molecules inside the bubble, thereby transporting heat without creating a temperature gradient. Also, according to kinetic theory, the root-mean-square velocity of vapor molecules within the bubble, $u_\mathrm{v}=\sqrt{3k_\mathrm{B}T_\mathrm{v}/m}=788$~\SI{}{\meter\per\second} when the vapor temperature $T_\mathrm{v}= 450$~K. This gives the ballistic heat transport time scale as $\sim\!h/u_\mathrm{v}= \mathcal{O}(0.1)$~ns. In our experiments, we find the torus film bubble to oscillate at time periods in the range of $\tau_{\omega}\sim \mathcal{O}(100)$ ns. Thus we can safely say that thermal equilibrium is established across the bubble interfaces during oscillations. On the other hand, the time scale of thermal relaxation in the water in the pore volume can be obtained as $\tau_\mathrm{w}\sim \alpha_\mathrm{th}/{D_\mathrm{p}}^2= \mathcal{O}(100)$~ns, where $\alpha_\mathrm{th}=k_\mathrm{w}/(\rho_\mathrm{w} c_\mathrm{p,w})$ is the thermal diffusivity of water. This is in the same order as $\tau_{\omega}$, while the lifetime of the stable torus bubble has been observed for more than \SI{100}{\micro\second}. This bodes well with our theory that the nanopore temperature distribution approaches steady state for the mean torus bubble size.  

However, in a real situation, the  heat transport through the bubble will be a combination of both diffusive and ballistic transport, since the Knudsen number of the bubble is close to the $\mathrm{Kn}=1$ limit. In our model, for the sake of simplicity, we have neglected continuum diffusive transport through the bubble, assuming that the heat flux in the liquid surrounding the bubble can be balanced by the ballistic heat flux without incurring a huge temperature drop within the bubble. Thus, the vapor temperature within the bubble remains uniform and the heat flux is balanced, ensuring thermal quasi-equilibrium of the bubble. It should be noted that when this assumption is adopted, the steady-state liquid temperature distribution within the pore is such that the corresponding steady-state current $I_\mathrm{b,a}$ [Fig.~S6(b)~\cite{supp}] obtained according to the second of Eqs.~\eqref{eq:ap7} in Appendix~\ref{appB} matches the experimental mean current for a steady-state torus bubble [Fig.~S17(c)~\cite{supp}], thus justifying our model.


\section{Bubble mechanical equilibrium}\label{appC}
For net equilibrium, the torus bubble should be in mechanical equilibrium in addition to thermal steady-state, which can be quantified by the Young--Laplace equation
\begin{equation}
P_\mathrm{v} = P_\mathrm{w} + \gamma K,
\label{eq:ap9}
\end{equation}
where $K$ is the curvature of the torus bubble surface and $\gamma$ is the temperature-dependent surface tension of water. Assuming chemical equilibrium between liquid and vapor, the vapor pressure of the bubble, ${P_\mathrm{v}}$, is related to the saturation vapor pressure $P_\mathrm{sat}$ and the vapor temperature ${T_\mathrm{v}}$ according to $P_\mathrm{v} = P_\mathrm{sat}\exp[(P_\mathrm{w}-P_\mathrm{sat})M_\mathrm{w}/(N_\mathrm{A}\rho_\mathrm{w} k_\mathrm{B}T_\mathrm{v})]$, where $P_\mathrm{sat}$ is evaluated at $T_\mathrm{v}^*$, assuming saturation conditions. Here, $M_\mathrm{w}$, $N_\mathrm{A}$, and $k_\mathrm{B}$ are the molecular weight of water, Avogadro's number, and Boltzmann's constant, respectively.

\section{Linear model for bubble oscillations}\label{appD}
To model the  thermal oscillation frequency of the torus bubble, we adopt and extend the linear and approximate model described in Hao and Prosperetti~\cite{hao1999dynamics} and originally proposed by Alekseev~\cite{alekseev1976nonsteady}. We assume that the bubble contains saturated vapor and is in thermal equilibrium with the surrounding liquid and the silicon nitride membrane surface on which it sits:  $T_\mathrm{v}=T_\mathrm{w}=T_\mathrm{m}$. Also, for small-amplitude oscillations in contact angle, the vapor pressure inside the bubble is assumed to vary along the saturation line. So, for a change in  bubble volume of $\Delta V_\mathrm{b}$ corresponding to a contact angle change of $\Delta \theta$, the expansion of bubble mass is given by
\begin{equation}
\Delta m_\mathrm{b} = \rho_\mathrm{v}\Delta V_\mathrm{b} + V_\mathrm{b}\Delta\rho_\mathrm{v}= \rho_\mathrm{v}\frac{dV_\mathrm{b}}{d\theta}\Delta \theta + V_\mathrm{b}\frac{d\rho_\mathrm{v}}{dT_\mathrm{v}}\Delta T_\mathrm{v},
\label{eq:ap10}
\end{equation}
where $\rho_\mathrm{v}$ is the saturation vapor density. When there is no external heating, the increase in bubble mass is caused by the evaporation of liquid on the interface, the latent heat of which causes a reduction in interface liquid temperature by $\Delta T_\mathrm{v}$. Therefore,
\begin{equation}
-S_\mathrm{b}\sqrt{\frac{2D_\mathrm{th}}{\omega}}\,\rho_\mathrm{w}c_\mathrm{w}\Delta T_\mathrm{w} = h_\mathrm{fg}\Delta m_\mathrm{b},
\label{eq:ap11}
\end{equation}
where $S_\mathrm{b}$ is the liquid/vapor surface area of the torus bubble. Here, $h_\mathrm{fg}$ and $D=k_\mathrm{w}/(\rho_\mathrm{w}c_\mathrm{w})$ are the latent heat and thermal diffusion coefficient, respectively. $k_\mathrm{w}$, $\rho_\mathrm{w}$, and $c_\mathrm{w}$ are the thermal conductivity, density, and specific heat capacity of water as  functions of the liquid temperature. For low-amplitude oscillations at low frequencies, it is safe to assume that the thermal equilibrium at the bubble/liquid and bubble/membrane interfaces is not lost: $\Delta T_\mathrm{v}=\Delta T_\mathrm{w}=\Delta T_\mathrm{m}$. The term $\sqrt{2D_\mathrm{th}/\omega}$~\cite{Prosperetti2017} represents the thermal relaxation length over which the temperature decrease  $\Delta T_\mathrm{w}$ is observed. Here, $\omega$ is the angular frequency of bubble oscillation. Now, in the presence of Joule heating, this energy balance equation needs to be modified to account for the additional Joule heat generation due to bubble expansion, $\Delta V_\mathrm{b}$, which tends to increase the temperature in the thermal relaxation length. Thus,
\begin{align}
\frac{d\hat{h}_\mathrm{c}}{dV_\mathrm{b}}S_\mathrm{b}&\sqrt{\frac{2D_\mathrm{th}}{\omega}}\Delta V_\mathrm{b} - h_\mathrm{tm}S_\mathrm{p}\Delta T_\mathrm{v}\nonumber\\[6pt]
={}& (h_\mathrm{fg}\rho_\mathrm{v} + P_\mathrm{v})\frac{\Delta V_\mathrm{b}}{\Delta t} \nonumber \\[3pt]
&+\left(S_\mathrm{b}\sqrt{\frac{2D_\mathrm{th}}{\omega}}\rho_\mathrm{w}c_\mathrm{w} + S_\mathrm{p}\sqrt{\frac{2D_\mathrm{m}}{\omega}}\rho_\mathrm{m}c_\mathrm{m} \right.\nonumber\\
&\hspace{18pt}\left. \vphantom{\sqrt{\frac{2D_\mathrm{th}}{\omega}}}+ h_\mathrm{fg} V_\mathrm{b}\frac{d\rho_\mathrm{v}}{dT_\mathrm{v}} \right)\!\frac{\Delta T_\mathrm{v}}{\Delta t},
\label{eq:ap12}
\end{align}
where $S_\mathrm{p}=2\pi R_\mathrm{p}L$ is the cylindrical nanopore surface area of the silicon nitride membrane and $\hat{h}_\mathrm{c}=H_\mathrm{c}/V_\mathrm{p,c}$ is the average Joule heat density over the cylindrical pore volume $V_\mathrm{p,c}$. $H_\mathrm{c}$ is calculated by simulating the steady-state Joule heat and temperature distributions in the presence of the torus bubble following the model described in Eqs.~\eqref{eq:ap4}--\eqref{eq:ap8}. $V_\mathrm{p,c}$ is the cylindrical pore volume, which extends for $1.66 L$ across the pore midline ($z=0$), enclosing the torus volume $V_\mathrm{b}$ [Figs.~\ref{fig:4}(a) and~\ref{fig:4}(b)]. $D_\mathrm{m} = k_\mathrm{m}/(\rho_\mathrm{m}c_\mathrm{m})$ is the thermal diffusivity of the silicon nitride membrane. As discussed in  Appendix~\ref{appB}, in the equilibrium bubble position, the net Joule heat production rate in $V_\mathrm{p,c}$ is balanced by the heat dissipation through the surrounding liquid and the bubble, thereby rendering a steady-state temperature distribution.
 
Using Joule heating simulations, we have calculated the increase in Joule heat density $\hat{h}_\mathrm{c}$ for different contact angles $\theta_\mathrm{e}$ of the torus bubble during pinned expansion [Fig.~S6(c)~\cite{supp}] and have found that for large  $\theta_\mathrm{e}$, when the bubble expands to constrict the pore volume, the Joule heat density increases in the cylindrical pore region: ${d\hat{h}_\mathrm{c}}/{dV_\mathrm{b}}>0$. This increase in heat generation is in turn balanced by a change in heat transfer through the membrane ($h_\mathrm{tm}S_\mathrm{p}\Delta T_\mathrm{v}$), latent heat consumption by the bubble ($h_\mathrm{fg}\rho_\mathrm{v} \Delta V_\mathrm{b}/\Delta t$), $P$--$V$ work done by the bubble, and sensible heat rise in the thermal relaxation length in the liquid and the silicon nitride. Here, the internal energy rise of the bubble is neglected because of its comparatively lower heat capacity than the liquid layers surrounding it. Thus, in short, Eq.~\eqref{eq:ap12} accounts for the energy balance of the bubble and the thermal boundary layers on its interfaces.

As the Eq.~\eqref{eq:ap12} contains time-derivative terms, the solution for $T_\mathrm{v}$ oscillations will not be in the same phase as that for $V_\mathrm{b}$ oscillations. We can write
\begin{equation}
\begin{aligned}
\theta &=\theta_\mathrm{e} - \theta_\mathrm{a}\exp(j\omega t), \\
V_\mathrm{b}  &=V_\mathrm{b,e} + V_\mathrm{b,a}\exp(j\omega t), \\
T_\mathrm{v} &=T_\mathrm{v,e} + T_\mathrm{v,a}\exp[j(\omega t + \delta)],
\end{aligned}
\label{eq:ap13}
\end{equation}
where $j=\sqrt{-1}$. Here, the oscillation amplitudes of contact angle, bubble volume, and bubble temperature are taken to be positive by convention, i.e., $\theta_\mathrm{a}>0$, $V_\mathrm{b,a}>0$, and $T_\mathrm{v,a}>0$. $\delta$ is the phase difference between volume expansion and vapor bubble temperature rise. Substituting Eq.~\eqref{eq:ap13} into Eq.~\eqref{eq:ap12}, we obtain
\begin{equation}
V_\mathrm{b,a}( B -j A ) = T_\mathrm{v,a}\exp(j\delta)( D +j C),
\label{eq:ap14}
\end{equation}
where 
\begin{gather*}
B =\frac{d\hat{h}_\mathrm{c}}{dV_\mathrm{b}}S_\mathrm{b}\sqrt{\frac{2D_\mathrm{th}}{\omega}}, \qquad A =\omega (h_\mathrm{fg}\rho_\mathrm{v} + P_\mathrm{v}), \\
 D =h_\mathrm{tm}S_\mathrm{p}, \qquad C =\omega\! \left(S_\mathrm{b}\sqrt{\frac{2D_\mathrm{th}}{\omega}}\,\rho_\mathrm{w}c_\mathrm{w} + h_\mathrm{fg} V_\mathrm{b}\!\frac{d\rho_\mathrm{v}}{dT_\mathrm{v}}\right).
\end{gather*}
Equating the real parts and imaginary parts of Eq.~\eqref{eq:ap14}, we arrive at
\begin{equation}
\begin{aligned}
\tan\delta &=\frac{ A  D + B  C }{ A  C - B  D },  \\[6pt]
\cos\delta &=\psi\!\left(\frac{ B  D - A  C }{ D^2+C^2}\right).
\end{aligned}
\label{eq:ap15}
\end{equation}
Because $ A >0$, $ B >0$, $ C >0$, and $ D >0$, for small positive values of $\delta$, $\tan \delta >0$ will be satisfied only when $ A  C > B  D $. Here $\psi={V_\mathrm{b,a}}/{T_\mathrm{v,a}}$. Now, the ratio of bubble volume amplitude and bubble temperature amplitude can be solved for from Eq.~\eqref{eq:ap15} as
\begin{equation}
\psi=-\sign(\delta)\left[\frac{D^2+C^2}{\sqrt{( A  C - B  D )^2 + ( A  D + B  C )^2}}\right].
\label{eq:ap16}
\end{equation}
It is interesting to note that when this ratio is negative, with increasing bubble volume, the vapor temperature will fall, and thus the vapor pressure inside the bubble is also expected to decrease along the saturation line $\Delta P_\mathrm{v}=(dP/dT)_\mathrm{sat}\Delta T_\mathrm{v}= {h_\mathrm{fg}\rho_\mathrm{v}}/{T_\mathrm{v}} \Delta T_\mathrm{v}$. Hence, a resisting force opposing bubble expansion, $F$, is developed:
\begin{equation}
F=S_\mathrm{b}\Delta(P_\mathrm{v} - P_\mathrm{w} - K\gamma ) =-\kappa \Delta R=-\kappa \frac{\Delta V_\mathrm{b}}{S_\mathrm{b}},  \label{eq:ap17a}
\end{equation}
i.e.,
\begin{equation}
S_\mathrm{b}\!\left(\frac{h_\mathrm{fg}\rho_\mathrm{v}}{T_\mathrm{v}}\Delta T_\mathrm{v} - \gamma \frac{dK}{dV_\mathrm{b}}\Delta V_\mathrm{b} -K\frac{d\gamma}{dT_\mathrm{v}}\Delta T_\mathrm{v} \right)
=-\kappa \frac{\Delta V_\mathrm{b}}{S_\mathrm{b}},
\label{eq:ap17b}
\end{equation}
where $\Delta R=\Delta V_\mathrm{b}/S_\mathrm{b}$ is the expansion in bubble size. Now, substituting $\Delta V_\mathrm{b} = V_\mathrm{b,a}\exp(j\omega t)$, $\Delta T_\mathrm{v} = T_\mathrm{v,a}\exp[j(\omega t + \delta)]$, and $\psi = V_\mathrm{b,a}/T_\mathrm{v,a}$ into Eq.~\eqref{eq:ap17b} and solving for the real part, we obtain
\begin{eqnarray}
\kappa = &&-{S_\text{b}}^2 \left[\frac{h_\text{fg}\rho_\text{v}}{\psi T_\text{v}}\cos(\delta) - \gamma \frac{dK}{dV_\text{b}} -K\frac{d\gamma}{dT_\text{v}}\frac{\cos(\delta)}{\psi} \right] \nonumber \\
\kappa = &&{{S_\text{b}}^2}\left[\eta_2 - \frac{h_\text{fg}\rho_\text{v}}{T_\text{v}}\frac{\cos(\delta)}{{\psi}}+\eta_1 \frac{\cos(\delta)}{{\psi}}\right]
\label{eq:ap171}
\end{eqnarray}
The imaginary part of Eq.~\eqref{eq:ap17b} would correspond to the damping term associated with temperature being out of phase with volume perturbation. Here, $\kappa$ is the mechanical stiffness of the torus bubble and $\gamma$ is the temperature-dependent surface tension of water. $\eta_1$ and $\eta_2$ are the curvature coefficients depending on the bubble curvature, which in turn depend on contact angle and pore size as shown in Fig.~S4. It should be noted that $\psi\eta_2$ is one order of magnitude higher than $\eta_1$, so variations in $\eta_1$ can be neglected for the sake of this discussion. It should be noted that we ignore the recoil force applied on the bubble interface due to evaporation/condensation. The recoil pressure, which is given by ${\vert \dot{m} \vert}^2(1/\rho_\mathrm{v}-1/\rho_\mathrm{w})\approx \rho_\mathrm{v}\omega^2\Delta R^2$, is proportional to the square of the bubble radius perturbation and thus can be neglected for small-amplitude oscillations.

\section{Reduced mass of pinned torus bubble}\label{appE}
Under the constraint of pinning, as a torus bubble with contact angle $\theta$ oscillates with a contact angle velocity  $\dot{\theta}$, the liquid surrounding the bubble also develops a velocity according to the continuity equation. Let the kinetic energy of the liquid be given by
\begin{equation}
E_\mathrm{k} = \tfrac{1}{2}\mathcal{M}(\theta)\dot{R}^2,
\label{eq:ap18}
\end{equation}
where $\mathcal{M}(\theta)$ is the reduced mass of the bubble~\cite{Prosperetti2017, Vincent2017} and $\dot{R}$ is the velocity of the bubble interface, which can be approximated by 
\[
\dot{R}=\frac{1}{S_\mathrm{b}}\frac{dV_\mathrm{b}}{d\theta}\dot{\theta}.
\]
For a spherical bubble of radius $R$, $\mathcal{M}$ can be determined analytically to be $\mathcal{M}=4\pi\rho_\mathrm{w}R^3$. However, for a torus bubble, we need to  calculate $\mathcal{M}$ numerically, by implementing potential flow theory~\cite{kundu2015fluid} on a finite element method (FEM) grid in MATLAB~\cite{MATLAB:2019}.

First, at a given $\theta$, ${dV_\mathrm{b}}/{d\theta}$ is calculated from the slope of $V_\mathrm{b}$ versus $\theta$ in Fig.~S3(f)~\cite{supp}. The continuity equation is invoked in the liquid surrounding the bubble in an axisymmetric coordinate system:
\begin{equation}
{\nabla}^2 \phi = 0,
\label{eq:ap19}
\end{equation}
where $\phi$ is the velocity potential. For a contact angle velocity  $\dot{\theta}$, we calculate the radial ($\Delta R/\Delta t$) and axial  ($\Delta Z/\Delta t$) velocities of the tracer points on the bubble surface [Fig.~S8(b)~\cite{supp}]. These velocities are interpolated on the mesh points of the bubble surface, which act as boundary conditions. A far-field boundary condition  $\phi=\phi_0$ is applied on the liquid boundary far from the bubble. On the remaining boundary surfaces, zero-flux boundary conditions are applied: $ d\phi/dn=0$, where $n$ is the normal direction to the surface. The solution for the velocity potential is used to obtain the radial liquid velocity $u_{r}= \partial\phi/\partial r$ and axial liquid velocity $u_{z}= \partial\phi/\partial z$, which are integrated over the mesh triangles to obtain the total liquid kinetic energy
\begin{equation}
E_\mathrm{k} = \int_V\rho_\mathrm{w}\tfrac{1}{2}(u_{r}^2 + u_{z}^2)\,dV. 
\label{eq:ap20}
\end{equation}
Here, the density of the liquid is taken as $\rho_\mathrm{w}=1000$~\SI{}{\kilogram\per\meter^3}. From Eqs.~\eqref{eq:ap18} and~\eqref{eq:ap20}, $\mathcal{M}(\theta)$ is obtained.

\bibliography{snbibliography_EDITED}

\end{document}